\documentclass[numberedappendix]{emulateapj}
\def\icarus{\ref@jnl{Icarus}}           % Icarus

\usepackage{color}
\usepackage{natbib}
\usepackage{booktabs}
\usepackage{amsmath}
\usepackage{subfigure}

\renewcommand{\arraystretch}{1.3}

\usepackage{parskip}

\slugcomment{Accepted for publication in The Astrophysical Journal}
\shorttitle{Magnetized planet outflows}
\shortauthors{Uribe et.al.}

\shorttitle{Type I Planet migration in a Magnetized Disk. I.}
\shortauthors{Uribe, Bans, \& K\"onigl}

\begin{document}

\title{Type I Planet Migration in a Magnetized Disk.\\ I. Effect of Large-Scale Vertical and Azimuthal Field Components}

\author{Ana Uribe, Alissa Bans,\altaffilmark{1} and Arieh K\"onigl}
\affil{Department of Astronomy and Astrophysics, University of Chicago, Chicago IL 60637, USA}
\altaffiltext{1}{present address: Astronomy Department, Adler Planetarium, Chicago, IL 60605, USA; auribe@oddjob.uchicago.edu}

%\email{auribe@oddjob.uchicago.edu}

\begin{abstract} 
We study the effects of a large-scale, ordered magnetic field in protoplanetary disks on Type I planet migration using a combination of numerical simulations in 2D and 3D and a linear perturbation analysis. Steady-state models of such disks require the inclusion of magnetic diffusivity. To make progress using ideal MHD, we focus on simplified field configurations, involving purely vertical ($B_z$) and azimuthal ($B_\varphi$) field components and a combination of the two. For each of the models we calculate the locations of the relevant resonances and of the turning points, which delineate the propagation regions of the MHD waves 
that transport angular momentum from the planet to the disk. We use both numerical and semianalytic methods to evaluate the cumulative back torque acting on the planet, and explore the effect of spatial gradients in the disk's physical variables on the results. We conclude that, under realistic (3D) circumstances, a large-scale magnetic field can slow down the inward migration that characterizes the underlying unmagnetized disk --- by up to a factor of $\sim 2$ when the magnetic pressure approaches the thermal pressure --- but it cannot reverse it. A previous inference that a pure-$B_\phi$ field whose amplitude decreases fast enough with radius leads to outward migration applies only in 2D. In fact, we find that, in 3D, a pure-$B_\phi$ disk undergoes a rapid transition to turbulence on account of a magnetorotational instability that is triggered by the planet-induced appearance of a weak $B_z$ component.
\end{abstract}


\section{INTRODUCTION}\label{intro}

Our current understanding of planets is that they form in the circumstellar disks from which their host stars originally accreted most of their mass, and that, particularly in the case of giant, largely gaseous, planets, this happens on timescales that are shorter than the dispersal times (typically a few million years) of these disks. The unanticipated finding of ``hot Jupiters,'' giant planets orbiting within $\sim 10$ stellar radii of their host stars, has indicated that such planets can move inward from their formation sites farther out in the protoplanetary disk. This outcome has been attributed to two possible mechanisms. The first involves a highly eccentric motion, induced by gravitational interaction with another planet (or planets) or with a companion star, and eventual circularization through tidal interaction with the central star \citep[e.g.,][]{RasioFord96,WuMurray03,WuLithwick11}. This process can take place after the disk has already dispersed. In contrast, the second mechanism, which envisions a slow inward drift while the planet maintains a quasi-circular orbit, requires the presence of a gaseous disk. In the latter picture, the planet exerts a gravitational perturbation on the surrounding gas, and the back reaction of this perturbation on the planet can generate a net torque that causes the planet to migrate (see, e.g., \citealt{LubowIda10}, \citealt{KleyNelson12}, \citealt{BaruteauMasset13}, and \citealt{BaruteauEtal13} for recent reviews). When a planet's mass is sufficiently small that one can neglect its effect on the disk structure, and if, in addition, its radial motion is slower than the effective viscous diffusion in the disk so that it can also be neglected, the planet is said to undergo Type~I migration. As was originally shown by \citet{GoldreichTremaine79}, the response of the gas in this case remains linear and can be represented by the effects of resonances: Lindblad resonances (LRs) that occur both inside and outside the corotation radius $r_{\rm c}$ (where the planet's angular velocity $\Omega_{\rm p}$ equals the angular velocity $\Omega$ of the gas), and a corotation resonance at $r_{\rm c}$. It turns out that, when the disk is isothermal, the LRs dominate the torque. At the LR locations, the planet exchanges angular momentum with the disk through the launching of rotation-modified sound waves that combine to produce a spiral wake. The waves from the outer region are excited in gas that rotates more slowly than the planet: they thus remove angular momentum from the planet and induce inward motion. The waves from the inner region have the opposite effect. In a Keplerian disk, the outer LR associated with a given azimuthal mode number lies closer to the planet than the corresponding inner LR. Therefore, if the density varies smoothly with radius, the interaction of the planet with the outer disk is stronger and the planet migrates inward.
The drift timescale for Type I migration is inversely proportional to the planet's mass $M_{\rm p}$ and is estimated to be shorter than the disk's lifetime even for an Earth-mass planet at a distance of $\sim 1\,$AU from a solar-mass star. The migration timescale stops decreasing with increasing $M_{\rm p}$ when the planet becomes massive enough to clear a gap in the disk through a nonlinear interaction, which implies that, so long as the local disk's mass remains $\gtrsim M_{\rm p}$, giant planets move with the radial velocity of the surrounding gas (Type~II migration). However, Type~I migration remains relevant to giant planet evolution in the context of the core-accretion model --- currently the favored scenario for giant-planet formation \citep[e.g.,][]{Armitage07,Rafikov09} --- in which such planets form through gas accretion onto comparatively small rocky cores after they exceed a threshold of a few~$M_\earth$. Within this modeling framework, the predicted rapid inward migration of the rocky embryos makes it difficult to account for the observed spatial distribution of giant planets \citep[e.g.,][]{IdaLin08a}.

The sign and magnitude of the net torque acting on a planet are determined by the difference between the opposing contributions from the regions inside and outside the planet's orbit as well as by the corotation torque from the co-orbital region, and can depend sensitively on the physical properties of the disk in the vicinity of the planet. The aforementioned ``rapid inward migration'' conundrum has motivated an exploration of more realistic disk models than the smooth, isothermal configuration that was originally studied. These investigations are still ongoing, although they have already yielded several promising clues. One notable result has been the finding that the corotation torque can become dominant when the disk is nonisothermal, and that the planet may, in fact, migrate away from the star if the disk has an outward-decreasing specific entropy and a sufficiently robust angular momentum transport mechanism to keep this torque unsaturated (see the above-cited reviews). Another intriguing possibility that has been discussed in the literature is the occurrence of disk inhomogeneities where the predicted direction of planet migration is reversed \citep[e.g.,][]{MassetEtal06,IdaLin08b,HasegawaPudritz11}. In addition, several studies have considered magnetic effects. Magnetic fields are expected to play a number of key roles in protoplanetary disks. They may contribute significantly to the angular momentum transport that enables gas to reach the central star --- either through magnetohydrodynamic (MHD) turbulence (likely induced by the magnetorotational instability [MRI]), which involves a disordered, small-scale field that transports angular momentum in the radial direction, or by means of an ordered, large-scale field, which can give rise to vertical transport (likely involving a centrifugally driven wind) through the disk surfaces (see, e.g., \citealt{Balbus11} and \citealt{KoniglSalmeron11} for recent reviews). In this connection, it has been proposed that the radial motion of small ($M_{\rm p}\lesssim M_\earth$) planets on scales of a few AUs could be dominated by stochastic migration induced by MRI turbulence \citep[e.g.,][]{Nelson05,JohnsonEtal06,AdamsBloch09,UribeEtal11}. Another relevant effect is the truncation of the disk by a sufficiently strong stellar magnetic field, which could help stall inward-migrating planets before they reach the central star \citep[][]{LinEtal96}. Finally, a large-scale, ordered magnetic field that permeates the disk and is well coupled to the gas can have a direct effect on planet migration by modifying the disk response to the planet's gravitational potential. This change is due to the fact, in the presence of such a field, perturbations are no longer associated with hydrodynamic (HD) sound waves but rather with either one of the three types of MHD waves: slow-magnetosonic (SMS), fast-magnetosonic (FMS), and Alfv\'en. 

The direct magnetic effect on planet migration was first considered in the semianalytic work of \citet{Terquem03}, who carried out a linear perturbation analysis for the case of an isothermal, 2D disk (i.e., one in which there are no variations along the rotation axis) that is permeated by a purely azimuthal magnetic field. The results of that study were confirmed numerically by \citet{FromangEtal05}. It was found that the perturbations induced by a planet embedded in such a disk give rise both to modified LRs, which occur at the locations where the azimuthal velocity of the gas relative to the planet equals the phase velocity of FMS waves in the azimuthal direction, and to so-called magnetic resonances, which occur at the locations where the relative azimuthal velocity matches the phase velocity of a SMS wave propagating along the field line.\footnote{As we note in Section~\ref{TPs}, the locations of the LRs actually correspond to turning points, rather than to genuine resonances, of the perturbation differential equation.} 
While the excitation at the modified LRs again gives rise to the launching of density waves that propagate away toward either the inner or outer disk, the waves excited at a magnetic resonance propagate in a restricted region around it. Since the magnetic resonances are closer to the planet than the LRs, their contribution to the torque dominates over that of the Lindblad torque if the magnetic field is large enough. It was also found that there is no corotation resonance in this case but that nevertheless the whole region around corotation (and not just narrow zones around the magnetic resonances) contributes to the total torque.\footnote{The \citet{Terquem03} 2D linearized perturbation analysis was carried out in the ideal-MHD limit. Under these conditions, it is conceptually not surprising that the corotation resonance was found not to be present in the magnetic case. This is because the corotation torque can be interpreted in terms of the drag exerted by particles that move on horseshoe orbits in the vicinity of the planet \citep[(e.g.,][]{Ward91}. In the ideal-MHD limit, the particles are ``frozen'' onto the magnetic field lines, and, for a purely azimuthal field, cannot cross the radius of the planet on a horseshoe orbit. It has, however, been found that, as the field weakens and the horseshoe motions (which act to suppress the magnetic resonances) are reestablished, the corotation torque acquires a new magnetic component that could potentially cause the planet to migrate outward \citep[e.g.,][]{GuiletEtal13}.} 
Similarly to the LRs, in this case the outer magnetic resonances induce inward migration whereas the inner ones have the opposite effect. This suggests that, if the field amplitude decreases sufficiently rapidly with radius $r$, inward migration could be slowed down or even reversed. This was indeed verified to be the situation for the 2D, purely azimuthal field case, where, for example, \citet{FromangEtal05} found (for a disk of uniform temperature and density in which the local thermal-to-magnetic pressure ratio is 4) that inward migration is reduced (resp., outward migration occurs) when the field scales as $r^{-1}$ (resp., $r^{-2}$).
 
The magnetic field configuration adopted in the aforementioned studies is not expected to arise naturally in protoplanetary disks. These disks likely form in the gravitational collapse of rotating molecular cloud cores that are threaded by a large-scale interstellar magnetic field. This picture is consistent with the hourglass magnetic field morphology exhibited by some of these cores, which can be attributed to the pinching and dragging of the field lines by the inflowing matter \citep[see, e.g.,][]{KoniglSalmeron11}. An open field of this type is conducive to the driving of disk winds, and, in its simplest representation, has an even symmetry about the disk midplane (corresponding to the radial and azimuthal field components changing sign across the midplane even as the vertical component remains nonzero there). In this case, the gas located in the vicinity of a planet at the midplane of the disk would be threaded by a purely vertical field rather than by a purely azimuthal field as was assumed in the above studies. As a first step toward generalizing the earlier results, \citet[][]{MutoEtal08} examined the effect of a purely vertical field on planet migration through analytic and numerical calculations. They showed that, in 2D, the only effect of the field is to change the thermal sound speed to the (higher) FMS wave speed, which leads to a reduction in the magnitude of the torque acting on the planet in comparison with the HD case. They also demonstrated that, in 3D, both SMS and Alfv\'en waves are excited in the disk. However, their study was carried out within the framework of the shearing-sheet approximation \citep[e.g.,][]{NarayanEtal87}, which does not distinguish between the regions outside and inside the planet's orbital radius and therefore does not allow a determination of the differential (net) torque acting on the planet. Furthermore, their numerical calculations in the 3D case were performed in the limit of a thermal-to-magnetic pressure ratio $\ll 1$, which is not relevant to protoplanetary disks (except perhaps near the boundary with the stellar magnetosphere).

In this paper and its companion (\citealt{BansEtal14}, hereafter Paper~II) we carry out a systematic study of the effects of a multi-component, ordered magnetic field on Type I disk migration using a linear perturbation analysis and ideal-MHD numerical simulations. As noted above, protoplanetary disks are expected to have a significant radial field component on account of the inward gravitational pull of the central star. In the presence of the strong azimuthal shear that characterizes Keplerian disks, this component would be rapidly wound up into a strong azimuthal component that would get out of equilibrium unless the disk is diffusive. While real disks likely possess extended regions that have a sufficiently large diffusivity to justify a quasi-equilibrium treatment, the numerical modeling of such disks, especially if vertical angular momentum transport involving a centrifugally driven wind is taken into account, is a challenging and costly endeavor. We defer simulating planet migration in this general case to a future publication and concentrate in this paper on gaining physical insights into this process through a series of model problems that can be considered within the framework of ideal MHD. Specifically, we investigate the cases of a purely vertical and a purely azimuthal field, as well as the case of a combined vertical and azimuthal field, in~2 and~3 dimensions. Our general linearization formalism is described in Paper~II, where it is illustrated with semianalytic results for a purely vertical field as well as for the case where an azimuthal field whose amplitude increases with height (from zero at the midplane) is present, alone or in conjunction with a vertical field (with the latter case enabling vertical angular momentum transport and thereby mimicking a real wind-driving disk). In Section~\ref{basicformulation} of this paper we briefly review the linearization procedure and then present analytic and semianalytic results for the case of a field with vertical and azimuthal components that are uniform with height but variable with radius. Our numerical simulations are presented and discussed in conjunction with the results of the linear analysis in Section~\ref{simulations}, with further discussion given in Section~\ref{discuss}. We summarize our findings in Section~\ref{conclude}.

\section{LINEAR ANALYSIS}\label{basicformulation}

In this section we perform a linear perturbation analysis to calculate the torque exerted by the disk on the planet. 
We expect the analytic results to complement the numerical simulations and to enable a fuller understanding of the disk migration process in the presence of a large-scale, ordered magnetic field. For the vertical and azimuthal fields we investigate, we present the locations where the density perturbation in the disk is divergent (the resonances) and also the locations where the planet can excite waves in the disk (the turning points). A detailed general formulation for calculating the effect of a planet on a magnetized disk is presented in Paper II.\footnote{In Paper~II we adopt the convention of previous semianalytic treatments and evaluate the torque that the planet exerts \emph{on the disk}, which has the opposite sign to the torque exerted \emph{by the disk} on the planet that we calculate in this paper.}  As we noted in Section~\ref{intro}, analytic studies were previously performed by \citet{Terquem03} for the case of a purely azimuthal field in 2D (see also \citealt{FromangEtal05}) and by \citet{MutoEtal08} for a purely vertical field in 2D and 3D. 

\subsection{Resonances}\label{resonances}

We first review the basic formulation, concentrating on the case where the field has only vertical and azimuthal components (in an inertial nonrotating cylindrical coordinate system $\{r,\varphi,z\}$). In the ideal-MHD limit, the conservation equations for momentum and mass, and the induction equation, take the form

\begin{equation} 
\rho \,  \Big [ \frac{\partial \textbf{v} }  { \partial t}  + ( \textbf{v} \cdot \nabla)  \textbf{v}  \Big ] = - \nabla  P  -  \rho \nabla \psi  + \frac{1}{4\pi}(\nabla \times \textbf{B}) \times \textbf{B}\; ,
\label{eom}
 \end{equation}

\begin{equation} 
  \frac{\partial \rho}{\partial t} + \nabla \cdot (\rho \textbf{v })  =0  \; , 
\label{con}
 \end{equation}

\begin{equation}  
\frac {\partial \textbf{B} } {\partial t} + \nabla \times (\textbf{v} \times \textbf{B} )  =0 \; ,
\label {ind}
\end{equation}
where $\rho$, $\textbf{v}$, and $P$ are the gas density, velocity, and pressure, respectively, $\psi$ is the gravitational potential, and $\textbf{B}$ is the magnetic field. The equilibrium velocity is taken to be $\textbf{v}_0=(0,r\Omega_0, 0)$, where $\Omega$ is the angular velocity and the subscript `0' denotes an equilibrium value. We assume that the disk is geometrically thin and rotates at nearly the Keplerian angular velocity; i.e., $\Omega_0 \approx \Omega_{\rm K}$. The equilibrium magnetic field is given by
$\textbf{B}_0 = (0,B_{\varphi \rm p}  r^{-q_{\varphi}}, B_{z\rm p} r ^{-q_{z}} )$, where we henceforth normalize $r$ by the radial location $r_{\rm p}$ of the planet (subscript `p'). Unless noted otherwise, we take all equilibrium quantities to be uniform in $z$. The presence of the planet induces perturbations in these variables through its gravitational potential, given by Equation~(\ref{eq:gpot}) below. Working in Fourier space in the $t$, $\hat{\varphi}$, and $\hat{z}$ directions, we write all perturbed disk variables as $\delta X =  X^\prime(r) e^ { i (  m \varphi +  k_{z} z - \omega t )}$, with $\omega \equiv m \Omega_{\rm p}$ 
and $m$ a nonnegative integer. The planet's angular velocity $\Omega_{\rm p}$ is equal to the Keplerian angular velocity at $r_{\rm p}$.

We define the Lagrangian displacement, $\boldsymbol{ \xi}$, in terms of the perturbed velocities as
\begin{equation}
v_{r}^\prime= i m \sigma \xi_{r}\ ,   
\label{vr'}
 \end{equation}

\begin{equation}
v_{\varphi}^\prime= i m \sigma \xi_{\varphi}  - r \xi_{r} \frac{\partial \Omega} { \partial r}    \label{vphi'}\ ,
\end{equation}

\begin{equation}
v_{z}^\prime= i m \sigma \xi_{z} \ , 
 \label{vz'} 
\end{equation} 
where $\sigma \equiv 
\Omega_0 - \Omega_{\rm p}$. Using $\textbf{B}' = \nabla \times \big ( \boldsymbol{\xi}  \times \textbf{B} \big) \label{B'CF} $ \citep{ChandraFermi53} and the above definitions for $ \boldsymbol{ \xi}$, the linearized induction equation becomes 

\begin{equation}
 B_{r}^\prime =  i k_{z} B_{z\rm p} r^{-q_{z}} \xi_{r} + imB_{\varphi \rm p} r^ {- ( q_{\varphi} +1) } \xi_{r} \; , 
 \label{Brp} 
\end{equation}

\begin{eqnarray} 
 B_{\varphi}^\prime  &=& i k_{z}B_{z\rm p} r^{-q_{z}} \xi_{\varphi} + q_{\varphi}B_{\varphi\rm p} r^ {-(q_{\varphi} +1)}\xi_{r}\nonumber \\
   &-&i k_{z} B_{\varphi\rm p} r^ {-q_{\varphi}} \xi_{z}  - B_{\varphi \rm p}  r^{-q_{\varphi}} \frac{\partial \xi_{r}}{\partial r} \ ,
  \label{Bpp}
  \end{eqnarray}

\begin{eqnarray} 
  B_{z}^\prime = &-& im B_{z\rm p} r^{-(q_{z}+1)}\xi_{\varphi} + (q_{z}-1)B_{z\rm p} r^{-(q_{z}+1)} \xi_{r}\nonumber \\
   &+& imB_{\varphi\rm p}r^{-(q_{\varphi}+1)} \xi_{z} -B_{z\rm p}  r^{-q_{z}} \frac{\partial \xi_{r}}{\partial r} \ .  
 \label{Bzp} 
\end{eqnarray}   

Linearizing Equations~\eqref{eom} -- \eqref{ind}, plugging into Equations~\eqref{Brp} -- \eqref{Bzp}, and rearranging the resulting terms, yields a second-order differential equation for $\xi_{r}$:

\begin{equation}
 A_{2} (r)\frac{\partial^{2} \xi_{r}}{\partial r ^{2}} + A_{1} (r) \frac{\partial \xi_{r}}{\partial r } + A_{0} (r) \xi_{r}  = S_{1} (r) \frac {\partial \psi^\prime}{\partial r} + S_{0} (r) \psi^\prime  
 \label{Diff} \ . 
\end{equation}
When the leading-order coefficient $A_{2}$ vanishes, the differential equation becomes singular: the locations where this happens represent \emph{resonances}. In the case of a field that has vertical and azimuthal components, the leading-order coefficient 
becomes\footnote{The other coefficients in Equation \eqref{Diff} cannot be written out in such a relatively simple form. All the functions that appear as coefficients in Equation \eqref{Diff} ($A_0$, $A_1$, $A_2$, $S_0$, and $S_1$) are, however, provided as online supplementary material in the form of both a Mathematica notebook and a PDF document.}
\begin{eqnarray}  
\label{A2Full}
A_{2} &=&  \left [ m^{2}\sigma^{2} - ( v_{{\rm A}z}k_{z} + v_{{\rm A}\varphi}k_{\varphi} )^{2} \right ]  \\
&\times& \left [( m^{2}\sigma^{2}(v_{\rm A}^2 + c^{2}) - c^{2}(v_{{\rm A}z}k_{z} + v_{{\rm A}\varphi}k_{\varphi})^{2} \right ] \nonumber \\
&\times&  [ - c^2 k^2 \left [ ( v_{{\rm A}z}k_{z} + v_{{\rm A}\varphi}k_{\varphi} )^2-m^2\sigma^2\right ] \nonumber \\
&+&  m^2 \sigma ^2 (v_{\rm A}^2 k^2-m^2 \sigma^2) ]^{-1} \ . \nonumber
\end{eqnarray}
Here $c$ is the isothermal speed of sound (i.e., $c^2=P/\rho$), and $v_{{\rm A}z}$, $v_{{\rm A}\varphi}$ are the Alfv\'{e}n speeds associated with the $\hat{z}$ and $\hat{\varphi}$ components of the field, respectively (i.e., $v_{{\rm A}z}^2= B_{oz}^{2}/4\pi\rho_0$, $v_{{\rm A}\varphi}^2=B_{0\varphi}^{2}/4\pi\rho_0$). For notational consistency we set $k_{\varphi}\equiv m/r$, and for the sake of brevity we use $v_{\rm A}^2 = v_{{\rm A}z}^2 + v_{{\rm A}\varphi}^2$ and $k^2 = k_z^2 + k_\varphi^2$. The numerator of Equation \eqref{A2Full} vanishes in two different cases.
For an arbitrary value of $m$, the first one applies when either $k_z \ne 0$ or both $v_{{\rm A}z}$ and  $v_{{\rm A}\varphi}$ are nonzero. It reads
\begin{equation} 
 m^{2}\sigma^{2} = (v_{{\rm A}z}k_{z} + v_{{\rm A}\varphi}k_{\varphi})^{2} \; ,
 \label{ARFull}
 \end{equation}
which indicates a resonance where the (Doppler shifted) forcing frequency $m\sigma$
equals the frequency of an Alfv\'{e}n wave. We therefore refer to this resonance as the  \emph{Alfv\'{e}n} resonance (AR). The second case is 
\begin{equation}
 m^{2}\sigma^{2}= \frac{c^{2}(v_{{\rm A}z}k_{z} + v_{{\rm A}\varphi}k_{\varphi})^{2}}{v_{{\rm A}z}^{2} + v_{{\rm A}\varphi}^{2} + c^{2}}  \; ,
  \label {MRFull}
\end{equation}
which indicates that there is another resonance where the forcing frequency matches that of a slow MHD wave; we refer to this second resonance as the \emph{magnetic} resonance (MR). These four resonances (inner/outer AR and inner/outer MR) of a $B_z+B_\varphi$ field configuration were previously discussed by \citet{FuLai11} in the context of black-hole accretion disks.

In the respective limits, Equation \eqref{A2Full} yields the pure-$B_z$ and pure-$B_\varphi$ cases that were considered in previous work. In particular, 
\begin{eqnarray}
\label{A2PureBz} 
&& A_{2} (B_{\varphi} \rightarrow 0) = \\
&&\frac{ (m^{2}\sigma^{2} - v_{{\rm A}z}^2k_{z}^{2}) (m^{2}\sigma^{2}(v_{{\rm A}z}^{2}  + c^{2}) - c^{2}v_{{\rm A}z}^2k_{z}^{2}) }{ - c^2 k^2 ( v_{{\rm A}z}^2k_{z}^2 -m^2\sigma
  ^2)+ m^2\sigma^2 \big( v_{{\rm A}z}^{2} k^2-m^2 \sigma^2 \big) }\ .  \nonumber 
\end{eqnarray}
The numerator of Equation \eqref{A2PureBz} vanishes at the Alfv\'en and magnetic resonances that were previously found for this case by \citet{MutoEtal08}. We note that these authors identified a third, ``vertical'' resonance from the vanishing of the denominator of Equation \eqref{A2PureBz} in the limit $k_\varphi \rightarrow 0$ (which corresponds to the WKB approximation employed in their derivation).
However, according to the definition given in Section~\ref{TPs}, this is technically a turning point and not a resonance. In the case of a pure-$B_\varphi$ field in the 2D limit  ($k_{z}=0$), Equation \eqref{A2Full} gives,
after factors of $(m^{2}\sigma^{2} -  v_{{\rm A}\varphi}^2k_{\varphi}^{2})$ are canceled from its numerator and denominator,
\begin{equation}   
A_{2} (B_{z}=0, k_{z}=0)= \frac{ m^{2}\sigma^{2}(  v_{{\rm A}\varphi}^{2} + c^{2}) - c^{2}v_{{\rm A}\varphi}^2k_{\varphi}^{2}}{   c^2  k_{\varphi}^2 -m^2 \sigma ^2 }\ . \label{A2PureBphi} 
\end{equation}
This result reproduces the MR identified for this case by \citet{Terquem03}. 
In the limit $k_z=0$, Equations \eqref{ARFull} and  \eqref{MRFull} imply that the locations of the AR and the MR are independent of $m$, and, for a geometrically thin disk, are given (to order $h$)  by
\begin{subequations}
\label{eq:r_AR_MR}
\begin{align}
\label{eq:r_AR}
r_{\rm AR} &= r_{\rm c}\pm \frac{2h_{\rm p}}{3\sqrt{\beta_{\varphi\rm p}}}\ , \\
\label{eq:r_MR}
r_{\rm MR} &= r_{\rm c}\pm \frac{2h_{\rm p}}{3\sqrt{1+ \beta_{\varphi \rm p}+\beta_{\varphi \rm p}/\beta_{z \rm p}}} \ , 
\end{align}
\end{subequations}
where $\beta$ is the ratio of the squares of the sound and Alfven speeds (i.e., $\beta_z = c^2/v_{{\rm A}z}^2$, $\beta_{\varphi}= c^{2}/ v_{{\rm A}\varphi}^{2}$) and $h\equiv c/(\Omega_{\rm K} r_{\rm p})$.
\footnote{In the limit of a thin disk, $h$ ($\ll 1$) is equal to the (normalized) tidal density scale height, which, for a vertically uniform magnetic field, is a good measure of the effective half-thickness of the disk.} For a purely azimuthal field, Equation \eqref{eq:r_MR} reproduces equation~(40) in \citet{Terquem03}. 

The locations of the resonances are in general obtained with respect to $r_{\rm c}$ \citep[e.g.,][]{Ward97}, so we are interested in tracking this radius. Using the radial component of Equation \eqref{eom} for a locally isothermal disk in which the midplane density and the sound speed scale with radius as $\rho_0(z=0)=\rho_{\rm p} r^{-a}$ and $c_0 = c_{\rm p}r^{-a_s}$, we obtain
\begin{eqnarray}
\label{eq:rcorot2}
1 &=& \frac{1}{r_{\rm c}^{2}} -
\frac{h_{\rm p}^2}{r_{\rm c}}\biggl((a+2a_s)r_{\rm c}^{-2a_s} + \\
 && r_{\rm c}^a \biggl[\frac{q_z}{\beta_{z\rm p}}r_{\rm c}^{-2q_z} +  \frac{(q_\varphi - 1)}{\beta_{\varphi \rm p}} r_{\rm c}^{-2q_\varphi}\biggr] \biggr) \nonumber,
\end{eqnarray} 
where the $\beta$ parameters are evaluated at the midplane. The dependence of $r_{\rm c}$ on $\beta_{\varphi \rm p}$ is illustrated in Figure~\ref{fig:rcorot}, using the disk models employed in Figure~\ref{fig:plot2} below.

\begin{figure}[ht]
\centering
\includegraphics[width=0.5\textwidth]{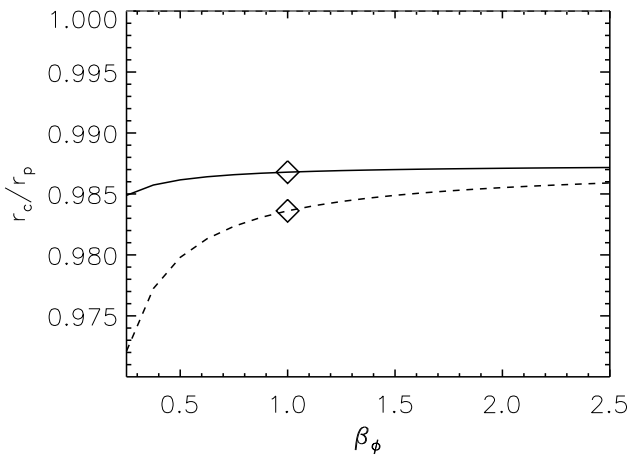}
\caption{Location of the normalized corotation radius (Equation \eqref{eq:rcorot2}) as a function of the azimuthal field-strength parameter $\beta_\varphi$ in a self-similar disk model ($a=3/2$, $a_{s}=1/2$, $q_\varphi=q_z=5/4$; note that the $\beta$ parameters are spatially constant in this model) for a pure-$B_\varphi$ field ($\beta_z \rightarrow \infty$; solid line) and for a $B_z+B_\varphi$ field configuration with $\beta_z = \beta_\varphi$ (dashed line). The diamond symbols mark the values of $r_{\rm c}$ for the two models in Figure~\ref{fig:plot2}.}
\label{fig:rcorot}
\end{figure}

\subsection{Resonances in a Strict-2D Limit} \label{pure2Dsec}

In order to compare the results of the analytic formulation with those obtained from our 2D numerical simulations (in which no vertical motion is allowed) we derive the leading-order coefficient of Equation \eqref{Diff} for the case where the condition $v_{z}^\prime=0$ is enforced. We note that this procedure is not fully self-consistent when both vertical and azimuthal equilibrium field components are present since the perturbed magnetic force term in Equation \eqref{eom} has a vertical component in this case that will induce vertical motion. We nevertheless proceed with this formulation since it is useful for interpreting the behavior of the 2D simulation results. 

Under the above assumption, the leading-order coefficient of Equation \eqref{Diff} becomes 

\begin{equation} 
 A_{2,2D} =\frac{ r^{2}\sigma^{2} \big (  v_{{\rm A}\varphi}^2 +  v_{{\rm A}z}^2 + c^{2} \big)  - v_{{\rm A}z}^2v_{{\rm A}\varphi}^2 - c^{2}v_{{\rm A}\varphi}^2}{v_{{\rm A}z}^2 + c^{2} - r^{2}\sigma^{2} }  \ . \label{A22D}
 \end{equation}

\noindent
This leads to the following condition for the MR: 

\begin{equation} 
 r^{2}\sigma^{2} =  \frac { v_{{\rm A}\varphi}^{2} ( c^{2} + v_{{\rm A}z}^2) }{ v_{{\rm A}\varphi}^2  +  v_{{\rm A}z}^2 + c^{2} }  
 \label{2Dres}  \ .
 \end{equation}

\noindent
This equation takes the exact same form as equation~(39) in \citet{Terquem03}, which gives the MR condition for a purely azimuthal field, if the sound speed, $c$, is replaced by the effective sound speed, $c_{\rm eff}=\sqrt{c^2 + v_{{\rm A}z}^2}$. Therefore, in this strict-2D limit, the effect of a vertical field component is just to increase the sound speed, which pushes the resonances and the turning points (see Section~\ref{TPs}) \emph{away} from $r_{\rm c}$. This behavior of a vertical field in this limit was first pointed out by \citet{MutoEtal08}.

Interestingly, the implications of Equation \eqref{2Dres} are the opposite of those inferred from the $k_{z}=0$ limit of Equation \eqref{MRFull}. In the latter case, the addition of a vertical field to an azimuthal field pushes the MRs \emph{closer} to the corotation radius (see Equation \eqref{eq:r_MR}). The turning-point locations also differ for these two cases (see Figure~\ref{2DTPpanel}). These results highlight the fact that a strictly 2D analysis may not capture the true response of the system. In the next subsection we return to the general case and consider the physical locations of the various resonances and of  their associated regions of wave propagation.

\subsection{Turning Points} \label{TPs} 

The locations where the solution to Equation \eqref{Diff} goes from wave-like to evanescent are known as the \emph{turning points}. At these radii the planet interacts most strongly with the disk. The size and radial position  of the regions of wave propagation are important predictors of the rate and direction of planet migration. In general, the closer the wave propagation regions are to the planet, the stronger will the disk--planet interaction be. To prevent confusion in the ensuing discussion, we remind the reader that, by this definition, the Lindblad ``resonances" are actually turning points \citep[see][]{GoldreichTremaine79}. 

We find the turning points by changing variables so that the homogeneous equation corresponding to Equation \eqref{Diff} has only second- and zeroth-order derivatives, and then solve numerically for the roots of the coefficient in front of the zeroth-order term --- this procedure is detailed in \citet{Terquem03} and is also described in Paper~II. In the examples that we exhibit, each field component has $\beta_{\rm p}=1$ (so that, in cases where both  $B_{\varphi\rm p}$ and $B_{z\rm p}$ are nonzero, the total field is characterized by $\beta_{\rm p}=0.5$).\footnote{Note, however, that the linear analysis results are based on a sound speed ($c_{\rm p}= 0.03$) that is slightly lower than the value ($c_{\rm p}=0.1$) used in the numerical simulations. The reason for this difference is that a comparatively small value of $c$ lowers the computational requirements on the integration of Equation \eqref{Diff} by reducing the size of the integration domain, but that a larger sound speed facilitates resolving the locations of the resonances in the simulations.} We henceforth also adopt the value $q_{\varphi}=1$ for the radial power-law exponent of the equilibrium azimuthal field component, which corresponds to a ``force free" configuration (no radial force associated with this field component; $(1/r)B_{\varphi}\partial (rB_{\varphi})/\partial r=0$). 

\begin{figure*}[ht!]
\begin{minipage}[b]{0.95\textwidth}
  \centering
  \subfigure[]{\includegraphics[height=.4\textwidth,width=.47\textwidth]{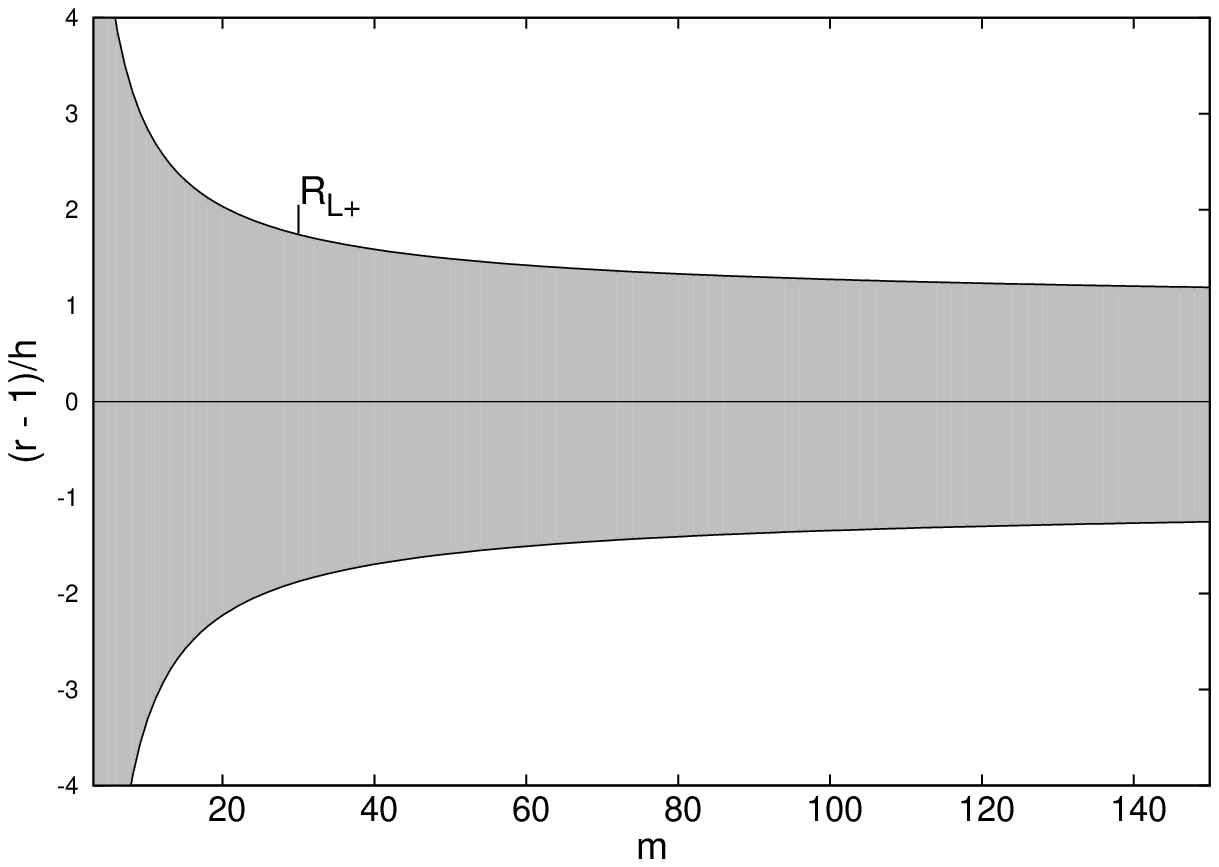}}\quad
  \subfigure[]{\includegraphics[height=.4\textwidth, width=.47\textwidth]{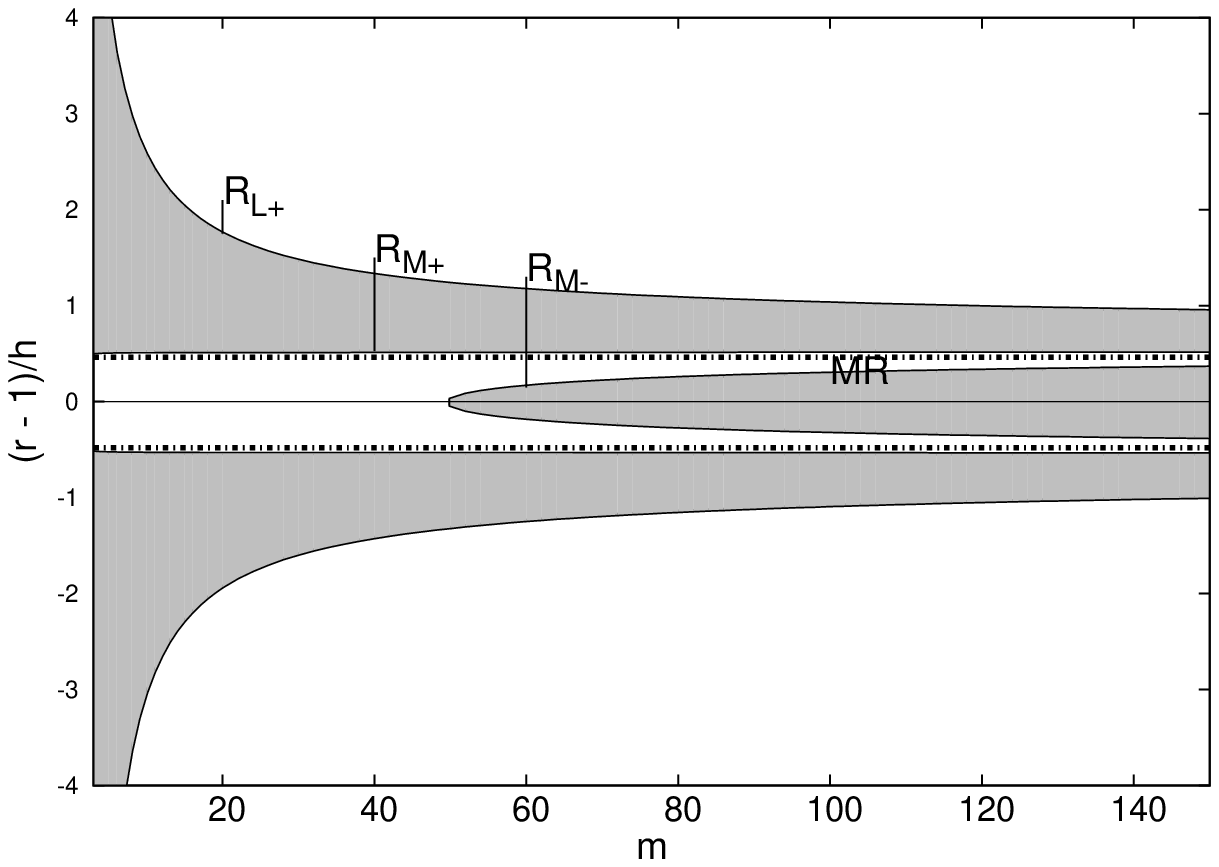}}\\
  \subfigure[]{\includegraphics[height=.4\textwidth, width=.47\textwidth]{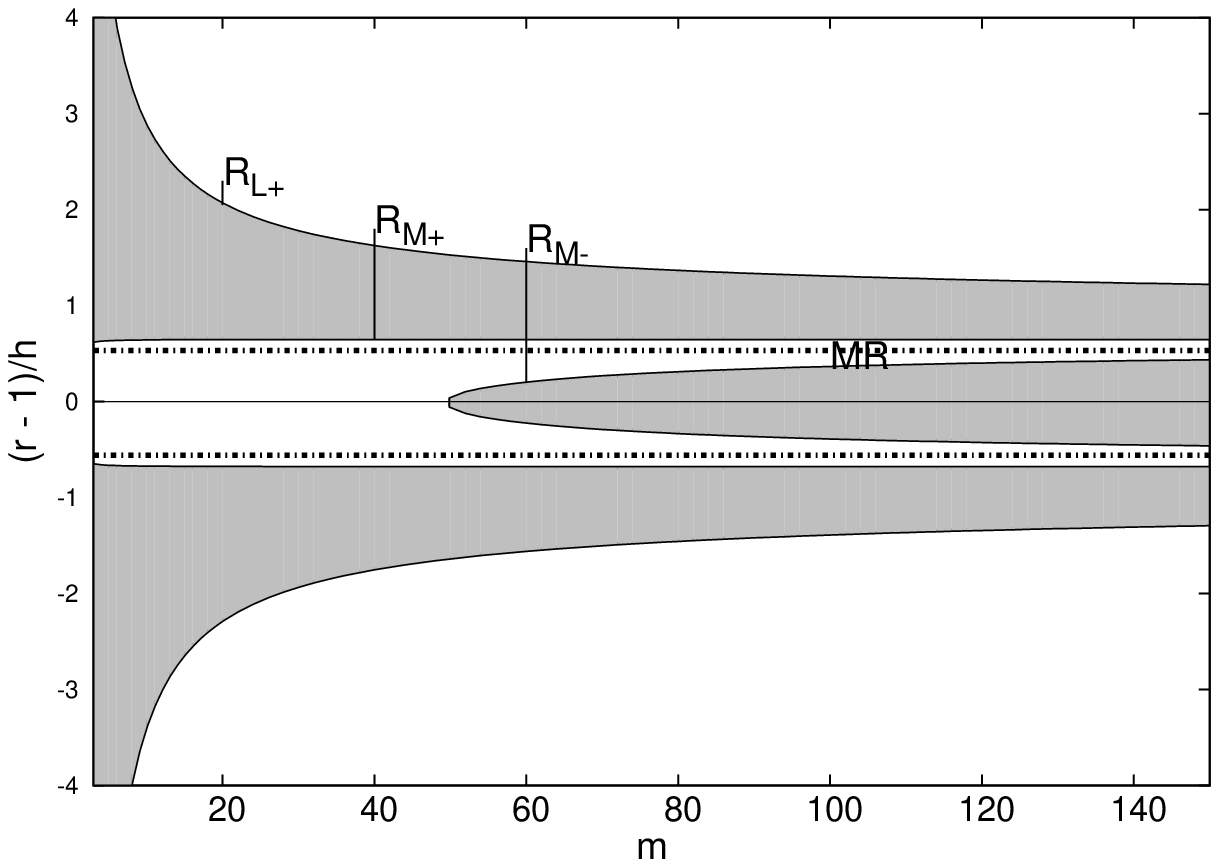}}\quad
  \subfigure[]{\includegraphics[height=.4\textwidth, width=.47\textwidth]{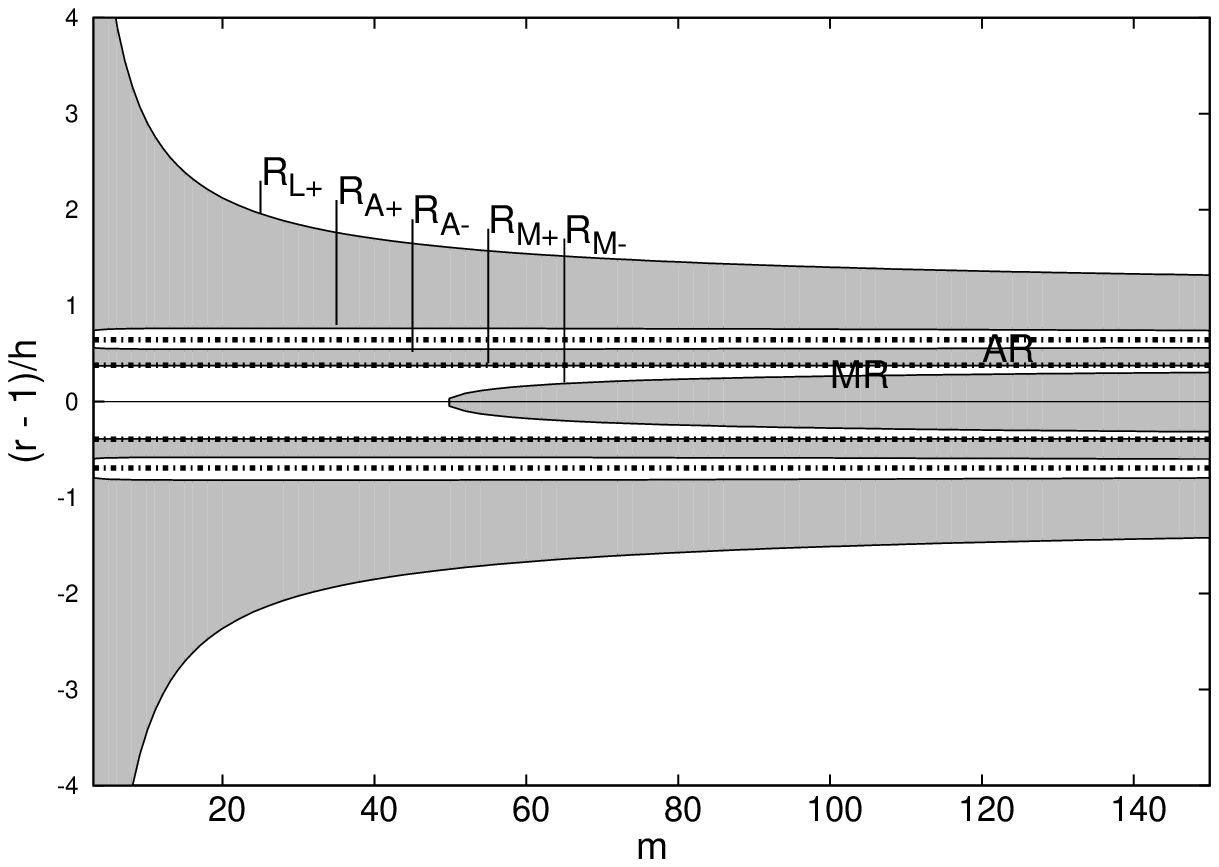}}\\
\end{minipage}
  \caption{Turning-point locations as a function of the azimuthal wavenumber $m$ in 2D for disks with the following field configurations: (a) pure $B_{z}$, (b) pure $B_{\varphi}$, (c)  $B_{z} + B_{\varphi}$ with $v_{z}^\prime =0$  (Section~\ref{pure2Dsec}), (d) $B_{z} +
  B_{\varphi}$ in the 2D limit ($k_{z}=0$) of the 3D model (Section~\ref{resonances}). The radial distance is normalized by the planetary orbital radius $r_{\rm p}$, and each field component is characterized by $\beta_{\rm p} = 1$. The shaded regions represent zones of wave evanescence, the dash-dotted lines mark the locations of the magnetic (MR) and Alfv\'en (AR) resonances, and the solid lines show the locations of the turning points (whose labels are described in the text).}
  \label{2DTPpanel}
\end{figure*}

Figure \ref{2DTPpanel} shows the turning points and regions of wave propagation for four different field configurations in the 2D limit. In the case of a purely vertical field (panel~(a)), there is only one set of turning points, corresponding to the effective Lindblad resonances.\footnote{The appellation ``effective'' is used to distinguish the turning points from the nominal Lindblad resonances, defined by $ m(\Omega - \Omega_{\rm p})=\pm \kappa$, where $\kappa$ is the disk's epicyclic frequency.} At high values of $m$, the locations of the effective Lindblad resonances deviate from those of the nominal ones due to a pressure effect, leading to a sharp cutoff in the torque \citep[e.g.,][]{GoldreichTremaine80,Artymowicz93}. A finite $B_{0z}$ component increases the total pressure (or, equivalently, the effective sound speed $c_{\rm eff} =\sqrt{ c^2 + v_{{\rm A}z}^{2}}$),  accentuating this effect and causing the positions of the high-$m$ turning points to shift farther away from the corresponding nominal resonances and thus remain farther away from the corotation radius (see Figure~II.3, where the label `II' henceforth refers to Paper~II). Therefore, as was previously noted by \citet{MutoEtal08}, the effect of a vertical field in 2D is to reduce the magnitude of the torque exerted on the planet. We label the modified (by the presence of a magnetic field) effective Lindblad resonances by $R_{L+}$.

In the case of the purely azimuthal field shown in panel~(b) of Figure \ref{2DTPpanel}, additional turning points appear on the two sides of the MR, corresponding to regions where SMS waves can propagate. We label the turning points interior and exterior to the MR by $R_{M-}$ and $R_{M+}$, respectively. In 2D, $R_{M-}$ exists everywhere only above some critical value of $m$ that depends on $\beta_\varphi$ and $h/r$ \citep[see][]{Terquem03}. 
The presence of additional regions of wave propagation, which are closer to the planet than the regions where the FMS waves associated with the effective Lindblad resonances propagate, suggests that the torque for this field configuration is stronger than for the pure-$B_z$ case. When the field strength is increased, all the turning points ($R_{M-}$, $R_{M+}$,  and $R_{L+}$) shift away from $r_{\rm c}$ (see figure~1 in \citealt{Terquem03}), which tends to decrease the torque. However, the amplitude of the SMS waves in the wave propagation regions surrounding the magnetic resonances goes up when the field gets stronger, which has the opposite effect. It turns out that the latter effect dominates, resulting in a larger torque for a stronger field in this case (in contrast with the behavior of a pure-$B_z$ disk).
  
As was shown in Section~\ref{pure2Dsec}, when one enforces the constraint $v_{z}^\prime=0$ for a disk with both $B_{\varphi}$ and $B_{z}$ field components, the magnetic resonance condition is the same as for a pure-$B_\varphi$ field except that $c$ is replaced by the (larger) effective sound speed $c_{\rm eff}$ (see Equation \eqref{2Dres}). The same effect is found in the behavior of the turning points for this field configuration, shown in panel~(c) of Figure~\ref{2DTPpanel}; all the turning points are shifted away from the corotation radius, as expected on the basis of such a substitution. This suggests that the disk would behave similarly to the pure-$B_\varphi$ case, but that the net torque would be reduced (resulting in slower inward migration). When the restriction $v_{z}^\prime = 0$ is removed (panel~(d)), one finds an additional set of turning points, with a corresponding new region of wave propagation. These turning points straddle the Alfv\'en resonance position, which, like the MR position in this case, is independent of $m$ (see Equation \eqref{eq:r_AR_MR}). We label the interior and exterior Alfv\'en turning points by $R_{A-}$ and $R_{A+}$, respectively. We also find that the MR, $R_{M-}$, and $R_{M+}$ loci now lie closer to the corotation radius than in the $v_{z}^\prime=0$ case. This fact, along with the appearance of new regions of wave propagation, lead to a stronger torque for the $v_{z}^\prime \ne 0$ setup, although the net torque for this case is still weaker than for the pure-$B_\varphi$ configuration (see left panel in Figure~\ref{fig:2DFFTorq} below). 
 
\begin{figure*}[ht!]
  \centering
  \subfigure[]{\includegraphics[height=.4\textwidth,width=.5\textwidth]{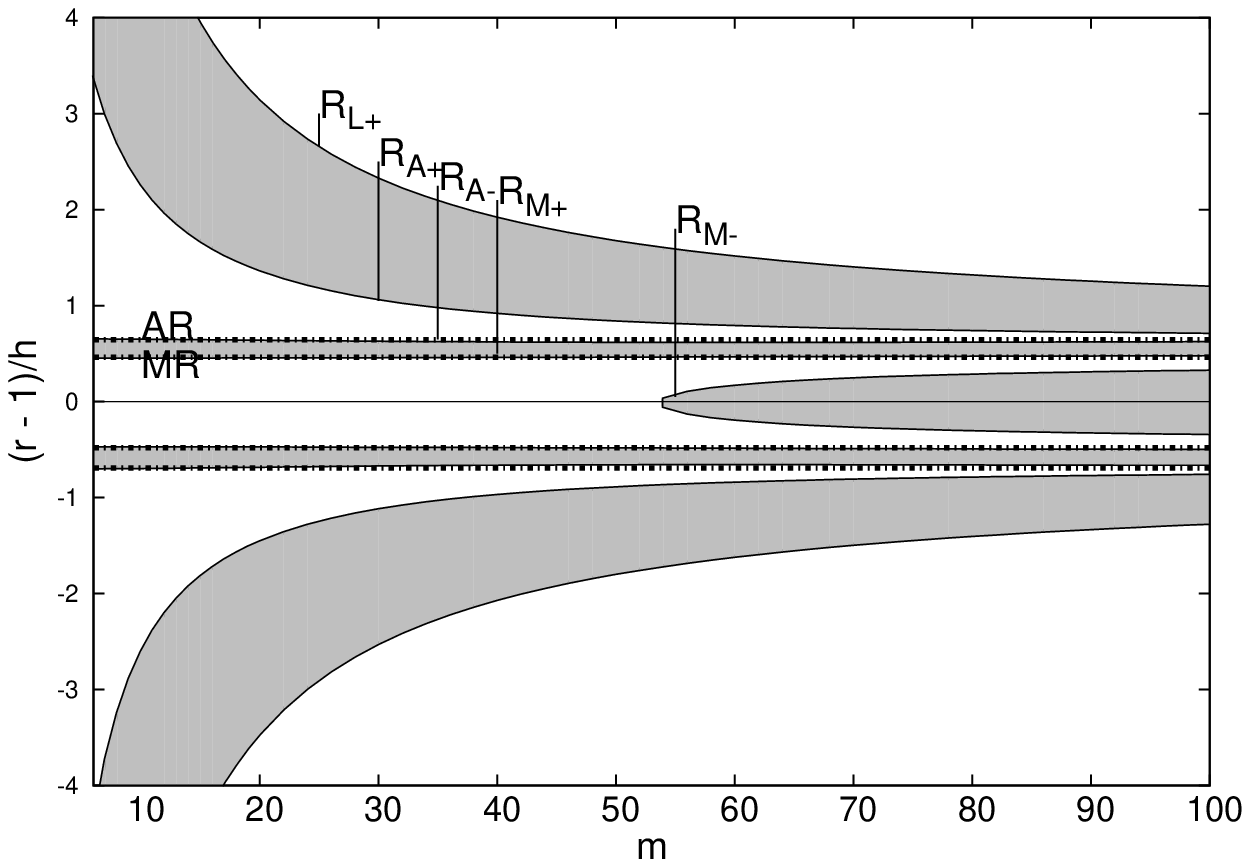}}\quad
  \subfigure[]{\includegraphics[height=.4\textwidth, width=.5\textwidth]{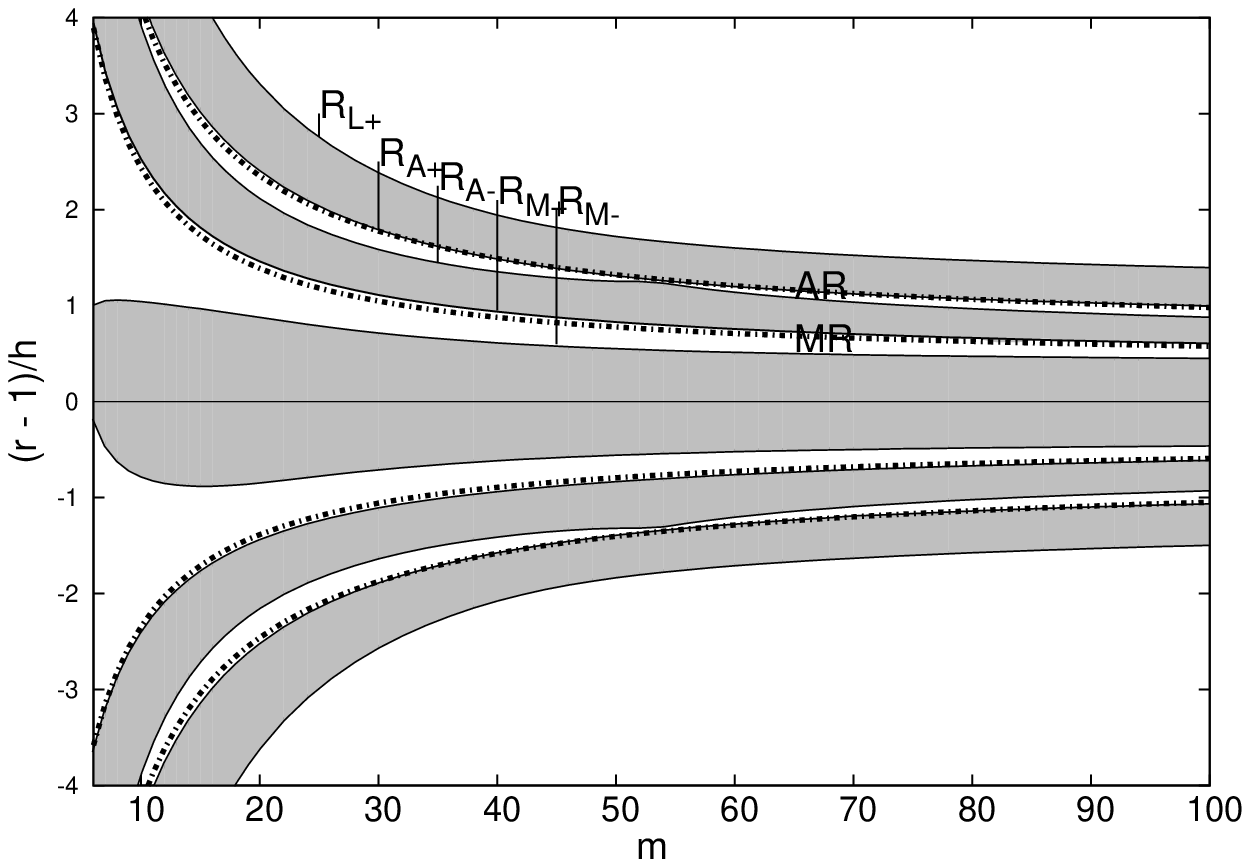}}\\
  \caption{Turning-point locations as a function of the azimuthal wavenumber $m$ in 3D for disks with the following field configurations: (a) pure  $B_{\varphi}$, (b)  $B_{\varphi} + B_{z}$, with each field component being characterized by $\beta_{\rm p} = 1$.The vertical wavenumber is given by $k_z h = 1.56$ in both panels, with $h = c/\Omega_{\rm p}$ being the nominal tidal scale height. Notation is the same as in Figure~\ref{2DTPpanel}.}
\label{3DTPpanel}
\end{figure*}

As is clear from our results in Section~\ref{resonances}, in 3D ($k_z \ne 0$) both the magnetic and the Alfv\'en resonances, as well as their associated turning points and wave-propagation regions (SMS and Alfv\'en, respectively), are present for either the pure-$B_z$, pure-$B_\varphi$, or their combined field configuration. The latter two cases are shown in Figure~\ref{3DTPpanel}. For the pure-$B_\varphi$ disk shown in panel~(a) of this figure, the innermost turning point ($R_{M-}$) is only present above a certain critical value of $m$, a behavior also seen in 2D (see panel~(b) of Figure~\ref{2DTPpanel}). However, we find that in 3D this value also has a slight dependence on the vertical wavenumber, becoming larger as $k_z$ is increased. The fact that wave propagation in the immediate vicinity of the planet is possible for low values of $m$ suggests that these azimuthal mode numbers would be the dominant contributors to the torque on the planet in this case. In contrast with this behavior, the innermost evanescent region extends over all values of $m$ for the $B_\varphi + B_z$ field configuration shown in panel~(b) of Figure~\ref{3DTPpanel}. This structure is evidently induced by the vertical field component, since it is similar to that of a 3D pure-$B_z$ field (see Figure~II.2). However, in the latter case the innermost turning points disappear for sufficiently low values of $v_{{\rm A}z}^{2}k_{z}^{2}$ (see Section~II.2.5 as well as \citealt{MutoEtal08}) --- we have confirmed that, when this product is small, the behavior of $R_{M-}$ resembles that of the innermost turning points in panel~(a) of Figure~\ref{3DTPpanel}.
 
\section{NUMERICAL SIMULATIONS}\label{simulations}

\subsection{Numerical Setup}\label{num_setup}

Our numerical simulations are performed with version 4.0 of the PLUTO code \citep{MignoneEtal07}. We employ the HLLC and HLLD approximate Riemann solvers for, respectively, HD and MHD computations, and use a second-order picewise parabolic spatial interpolation method and a second-order Runge-Kuta solver for time integrations. The Constrained Transport method is adopted in the MHD simulations to preserve a divergence-free magnetic field. Our calculations are carried out in the ideal regime (no explicit viscosity or resistivity). We use polar coordinates $\mathbf{r}=(r,\varphi)$ in the 2D simulations, and cylindrical coordinates $\mathbf{r}=(r,\varphi,z)$ in 3D.

The planet is assumed to move in a fixed, circular orbit, and is included in the simulation as an extra gravitational potential felt by the circumstellar disk. This potential has the form
\begin{equation}
\psi^\prime (\mathbf{r}) = -\frac{GM_{\rm p}}{(|\mathbf{r}-\mathbf{r}_{\rm p}|^2 +
\epsilon^{2})^{1/2}}\ ,
\label{eq:gpot}
\end{equation}
where $G$ is the gravitational constant and $\epsilon$ is a softening parameter (whose choice is discussed in Appendix~\ref{appA}). The planet's mass is taken to be a fraction $1.5\times 10^{-5}$ of the stellar mass $M_*$, and, using the same normalization as in Section~\ref{basicformulation}, the magnitude $r_{\rm p}$ of its orbital radius is set equal to 1. We carry out the simulations in the frame of reference of the star, neglecting the (small) stellar motion around the center of mass of the star--planet system.
For the 2D simulations, the planet is assumed move in the $x-y$ plane, and the computational domain is given by $r\in[0.4,2.5]$ and $\varphi \in [0,2\pi ]$. In this case only the radial and azimuthal equations are solved, and all vertical gradients ($\partial/\partial z$) are taken to be zero. In 2D, the number of cells in the computational domain is $(N_{r},N_{\varphi})=(512,1024)$ and the radial resolution is $\Delta r=0.004$ (approximately 1/10 of the radial distance from the planet to the MRs).
For the 3D calculations, the radial and azimuthal domains are the same, whereas the vertical domain is given by $z \in [-0.13,0.13]$. In 3D, the number of cells in the computational domain is $(N_{r},N_{z},N_{\varphi})=(512,64,512)$ and the vertical resolution is equal to the radial one ($\Delta z=0.004$). We calculate the $z$ component of the specific torque $\mathbf{\Gamma}$ exerted by the disk on the planet (hereafter the \emph{torque}) by integrating over the volume $V$ of the entire computational domain,
\begin{equation}
\Gamma_{z} = G \int \rho(\mathbf{r})\;
\frac{(\mathbf{r}_{\rm p}\times\mathbf{r})_{z}}{(|\mathbf{r}-\mathbf{r}_{\rm p}|^{2} + \epsilon^{2})^{3/2}}\; dV,
\end{equation}
where
$(\mathbf{r}_{\rm p}\times\mathbf{r})_{z}=(\mathbf{r}_{\rm p}\times\mathbf{r})\cdot \hat{e}_{z}$, and $\hat{e}_{z}$ is the Cartesian unit vector in the $z$ direction.

We refer to the running time average of the specific torque as the \emph{cumulative} torque. Torque is given in units of $r^2\Omega_{\rm K}^{2}$, evaluated at $r_{\rm p}$.

\vspace{0.07in}
\noindent
\textbf{Initial Conditions}

\vspace{0.07in}

The initial density profile of the disk is given by $\rho_0 = \rho_{\rm p}r^{-a}$. The adopted equation of state of the disk gas is $P=\rho c^{2}$, where $c$ is the isothermal sound speed (which we also take to have a power-law dependence on radius: $c = c_{\rm p}r^{-a_{s}}$). The tidal scale height is normalized by $h_{\rm p} =0.1$. 

Continuing to follow our analytic formulation, we take the initial field to be either purely azimuthal $\mathbf{B}_0=(0,B_{\varphi \rm p}r^{-q_\varphi},0)$, purely vertical $\mathbf{B}_0=(0,0,B_{z \rm p}r^{-q_z})$, or with both components $\mathbf{B}_0=(0,B_{\varphi \rm p}r^{-q_\varphi},B_{z\rm p}r^{q_z})$. We again parametrize the midplane field amplitudes by the $\beta$ parameters: $B_{\varphi p}= \sqrt{\mu P_{\rm p}/\beta_{\varphi \rm p}}$,  $B_{z\rm p}= \sqrt{\mu P_{\rm p}/\beta_{z\rm p}}$, where $\mu \equiv 4\pi$, and take $\beta_{\varphi \rm p}=\beta_{z\rm p}=1$ as the default values. (The motivation for these choices is that, in a well-coupled, wind-driving disk, $\beta_z$ decreases with height from a value  $\gtrsim 1$ at the midplane, whereas $\beta_\varphi$, which formally diverges at the midplane for an even field symmetry, typically satisfies $\beta_\varphi/\beta_z \lesssim 10$ at the disk surface; e.g., \citealt{WardleKonigl93}.) 

\vspace{0.07in}
\noindent
\textbf{Units}

\vspace{0.07in}

The density is given in physical units of $\rho_{\rm unit}=10^{-12}\; {\rm g\; cm}^{-3}$,
distances in units of $r_{\rm unit}=1\;$AU, and velocities in units
of $v_{\rm unit}=\sqrt{GM_\sun/r_{unit}}$. Magnetic fields are given in units of
$B_{\rm unit}=\sqrt{\mu\rho_{\rm unit} v_{\rm unit}^{2}}$. 

\vspace{0.07in}
\noindent
\textbf{Boundary conditions}

\vspace{0.07in}

For the 2D simulations, the inner and outer radial boundaries have zero-gradient boundary conditions for all physical quantities. We impose $v_r<0$ at the inner radial boundary and $v_{r}>0$ at the outer one. To prevent the development of high-frequency oscillations in the magnetic field, we apply the wave-damping procedure described in \citet{FromangEtal05}, which we implement by damping radial and azimuthal velocity perturbations near the radial boundaries ($r<0.55$ and $r>2.0$) with an exponential function. The azimuthal boundary has a periodic boundary condition. 

For the 3D simulations, we use the same boundary conditions as in 2D in the radial direction, but we apply no damping to velocity perturbations during the initial adjustment of the disk to a steady state. At the vertical boundaries, we implement the zero-gradient condition for all physical quantities whereas at the lower and upper $z$ boundaries we impose $v_{z}<0$ and $v_{z}>0$, respectively.

\subsection{Pure-$B_z$ Field in a 2D, Self-Similar Disk}\label{2Dsims_bz}

\begin{figure*}[t!]
\begin{minipage}[b]{0.24\textwidth}
   \includegraphics[width=\textwidth]{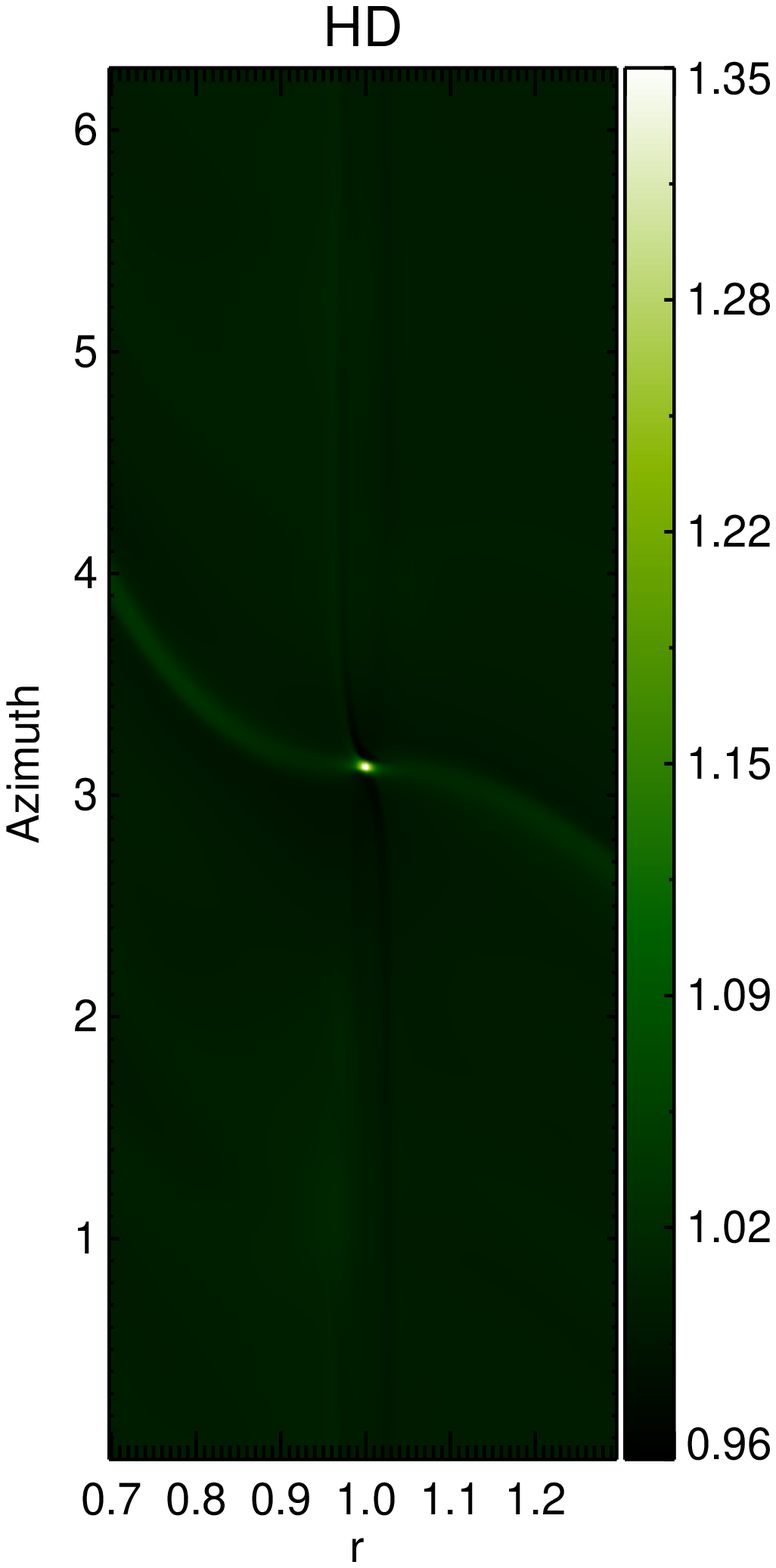}
\end{minipage}
\begin{minipage}[b]{0.24\textwidth}
   \includegraphics[width=\textwidth]{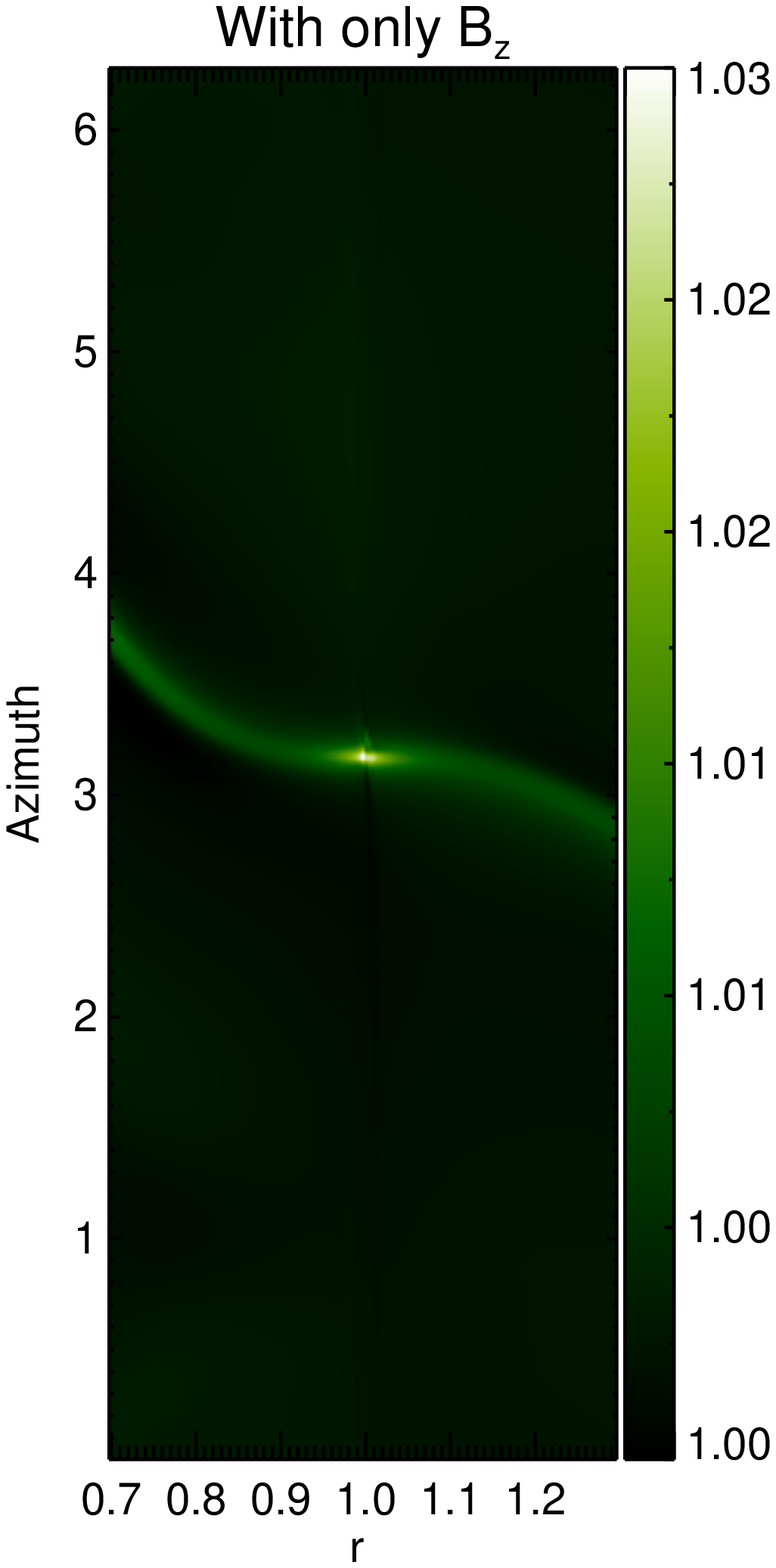}
\end{minipage}
\begin{minipage}[b]{0.24\textwidth}
   \includegraphics[width=\textwidth]{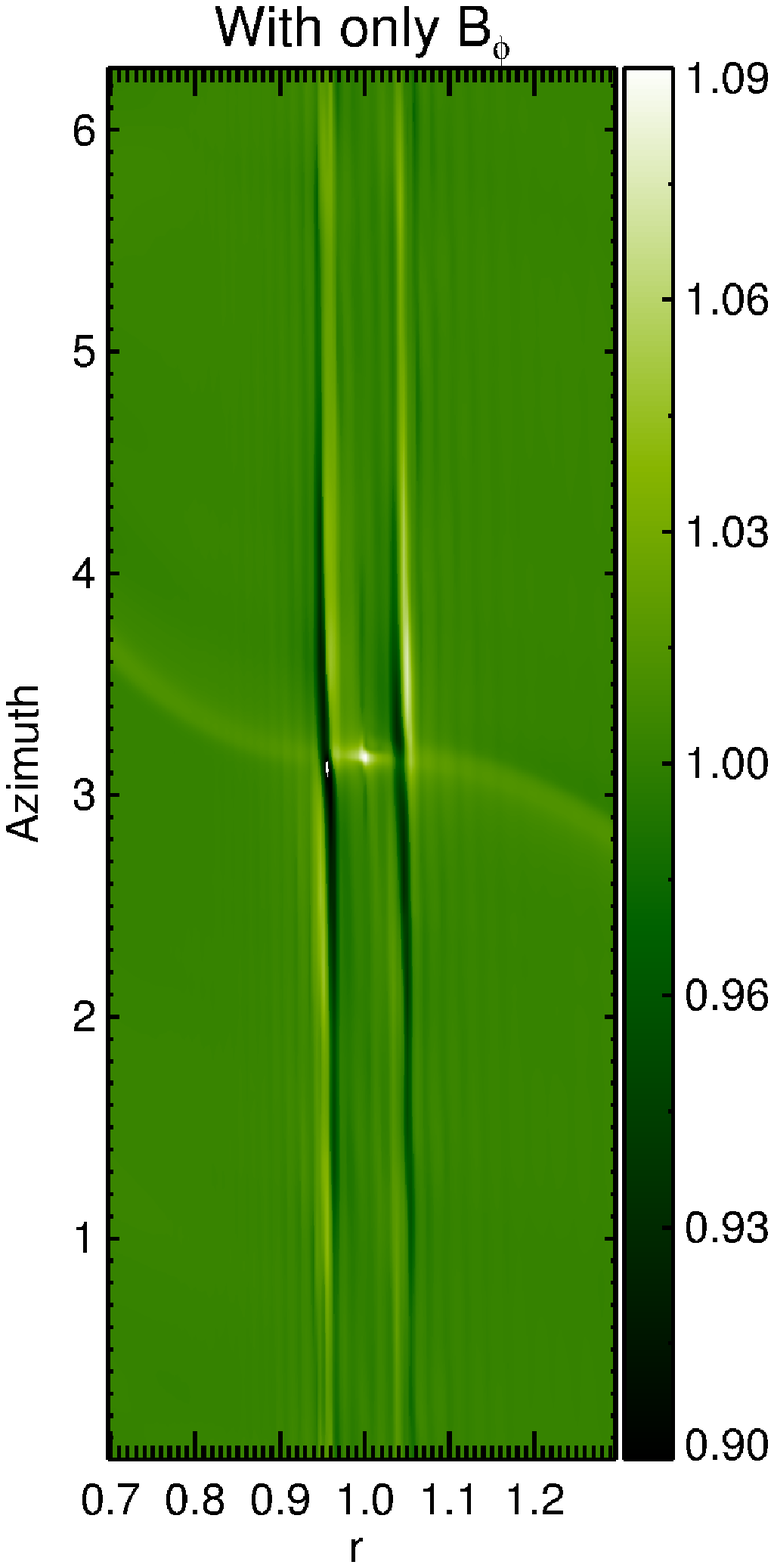}
\end{minipage}
\begin{minipage}[b]{0.24\textwidth}
   \includegraphics[width=\textwidth]{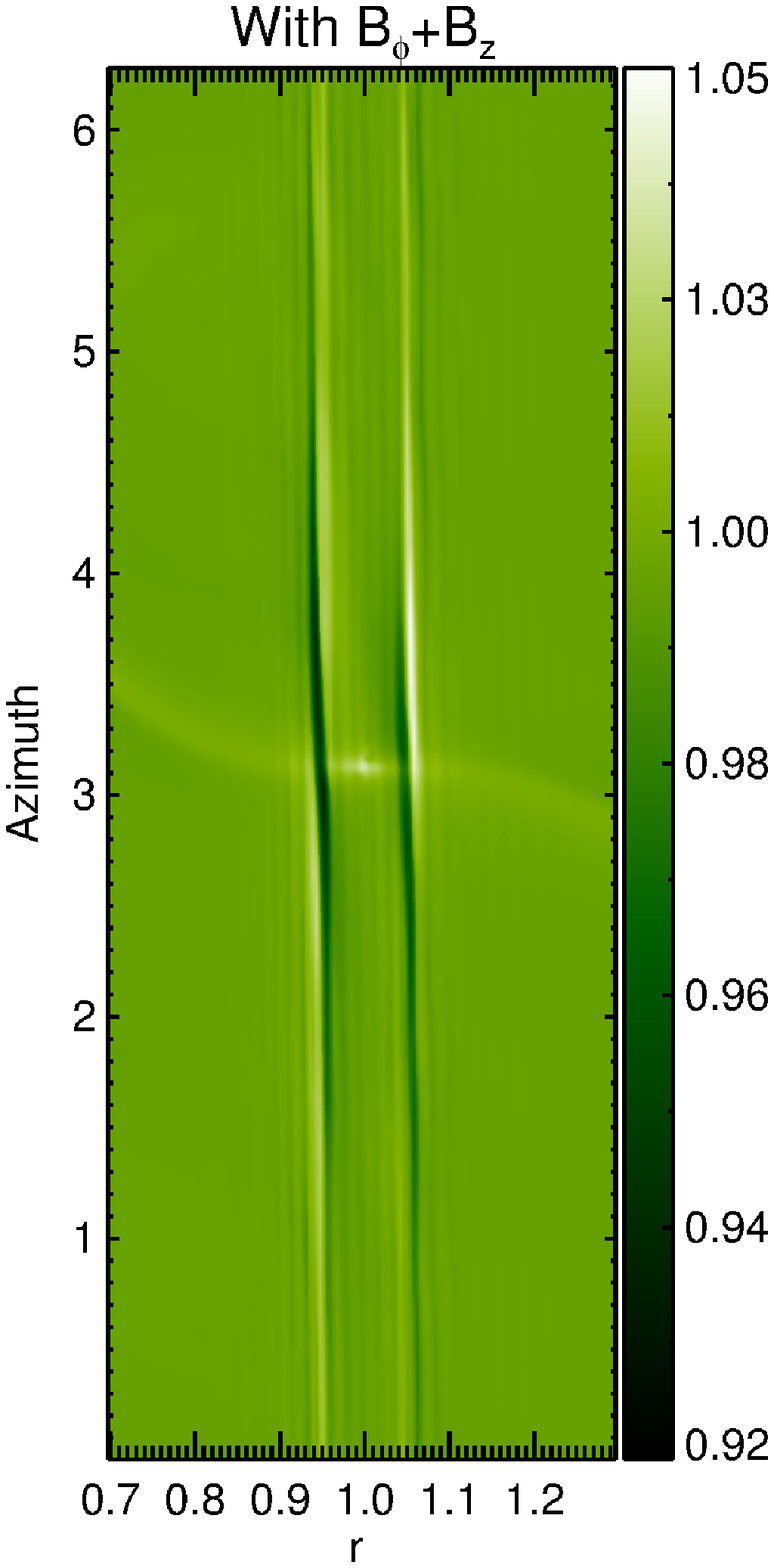}
\end{minipage}
\caption{Density in a uniform, 2D disk from simulations representing (from left to right) the HD (unmagnetized) case, a pure-$B_{z}$ field, a pure-$B_{\varphi}$ field, and a $B_{z} + B_{\varphi}$ configuration. The planet is located at $r=1.0$ and $\varphi=\pi$. All models show the spiral wake induced by the Lindblad resonances, and disks with an azimuthal field component also exhibit magnetic resonances (manifested as the density perturbations that extend along the $\varphi$ axis) both interior and exterior to the planet's location. Each field component is characterized by $\beta_{\rm p}=1$. }\label{fig:rphi_plots_rho}
\end{figure*}

\begin{figure}[t!]
\begin{minipage}[b]{\textwidth}
   \includegraphics[width=0.5\textwidth]{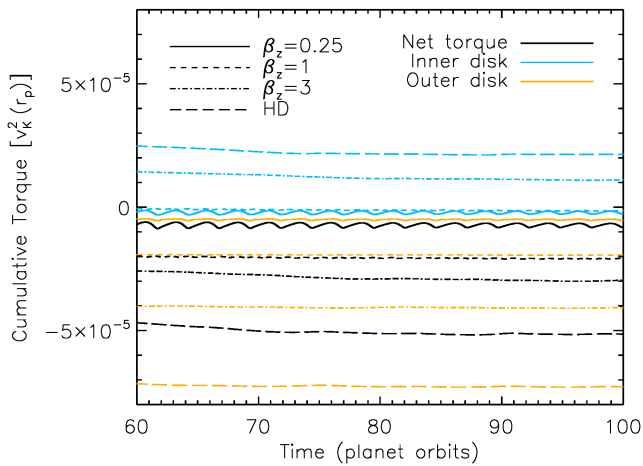}
\end{minipage}
\caption{Cumulative torques on the planet in a 2D disk as a function of time for a pure-$B_{z}$ field configuration. The different lines, specified in the legend, present the contributions from the regions interior and exterior to the planet, as well as the net torque, for different values of the magnetic field-strength parameter $\beta_z$. For these runs, the values of the power-law exponents for the radial dependence of $\rho$, $c$, and $B_z$ are $a=3/2$, $a_{s}=1/2$, and $q_{z}=5/4$, respectively, corresponding to a self-similar disk.}\label{fig:tors_bz}
\end{figure}

The first case we consider is that of a purely vertical field, which represents the midplane structure of a disk that is threaded by open magnetic field lines and possesses an even field symmetry with respect to the $z=0$ plane (the generic model for a wind-driving disk). This case was previously studied numerically by \citet{MutoEtal08}, albeit under several restrictive simplifications (see Section~\ref{intro}). 
For our simulations we adopt $a=3/2, a_{s}=1/2$ ,and $q_{z}=5/4$ --- this choice of radial gradients is motivated by the scalings obtained from radially self-similar models of magnetized wind-driving disks \citep[e.g.,][]{BP82,Koenigl89,Teitler11}.

In the 2D limit, the magnetic term in the radial momentum equation ($-(B_z/\mu)\partial B_{z}/\partial r$) acts simply as an extra radial pressure force, so we expect the disk to behave as if it had a larger sound speed. As we discussed in Section~\ref{basicformulation}, the effect of this on the effective LRs is to shift the locations of the turning points by a factor $(1+1/\beta_{z})^{1/2}$ away from $r_{\rm c}$. This leads to a weaker coupling of the disk to the gravitational potential of the planet, which reduces the amplitude of the induced perturbations at these locations. Consequently, there is a reduction in the magnitude of the torque acting on the planet. In addition, for the adopted radial scalings, Equation \eqref{eq:rcorot2} shows that the corotation point is shifted inward (so that $r_{\rm c}<1$), the shift being larger for a stronger field (i.e., a lower $\beta_{z}$). The effect of this inward shift is to increase the magnitude of the torque exerted by the outer disk while reducing the contribution (and potentially even changing the sign) of the torque exerted by the inner disk. 
The net torque on the planet is affected by both of these factors, the shift of the turning points and the shift of $r_{\rm c}$, but in general the shift of the effective LRs away from $r_{\rm c}$ will be the dominant effect (resulting in a lower torque from both the inner and outer disk regions).

Our simulations confirm these predictions. Figure~\ref{fig:rphi_plots_rho} (first 2 panels) shows the final (steady state) density structure in the $r-\varphi$ plane for an HD and a pure-$B_z$ disk. It  demonstrates that the same basic spiral wake structure, arising from the constructive interference of waves excited at the LRs, characterize both of these models. Figure~\ref{fig:tors_bz}, in turn, shows the cumulative torques exerted on the planet as a function of time for the fiducial pure-$B_z$ case ($\beta_z=1$) as well as for disk models having $\beta_{z}=0.25$, 3, and $\infty$ (the HD limit). We present the separate contributions of the regions interior and exterior to the planet as well as their sum. The ability to evaluate the differential torque distinguishes our simulations from those of \citet{MutoEtal08}, which were carried out under the shearing-sheet approximation. We find that the planet migrates inward, as in both the HD case and the ``pure $B_z$ in a uniform disk'' case studied semianalytically in Paper~II (see Figure~II.5). Our numerical results verify that the magnitude of the net torque decreases (implying slower inward migration) with increasing field strength. It is seen that the torque exerted by the inner disk is almost completely cut off for $\beta_z \leq 1$. In fact, for these low values of $\beta_z$ it even contributes (albeit very weakly) with the same sign ($<0$) as the outer disk. This seemingly paradoxical behavior is a reflection of the fact that the effect of the shifted LRs, which embody the disk's response to the planet's perturbing potential, is determined by their location relative to $r_{\rm c}$ rather than $r_{\rm p}$ (see Section~\ref{resonances}), and that for the low-$\beta_z$ models the region between $r_{\rm c}$ and $r_{\rm p}$ becomes large enough to dominate 
the ``inner'' torque.\footnote{As can be seen from Equation \eqref{eq:rcorot2}, the shift in $r_{\rm c}$ depends not just on the value 
of $\beta_{z \rm p}$ but also on the radial power-law exponents of the density, sound speed, and magnetic field. The corotation radius is shifted to a smaller radius than $r_{\rm p}$ when these quantities \emph{decrease} with $r$.} In a real wind-driving disk, however, the magnetic pressure would not exceed the thermal pressure at the midplane \citep[e.g.][]{KoniglSalmeron11}. For the lowest value of $\beta_z$ that we simulate (0.25), the net torque is reduced to $\sim 40\%$ of the HD value.

\subsection{$B_z+B_\varphi$ Field Configuration in 2D}\label{2Dsims_combined}

We now generalize the pure-$B_z$ disk model considered in the preceding subsection to the case where the field has both vertical and azimuthal components, which we take to be of equal magnitude. This corresponds to the field morphology near the surface of a wind-driving disk (although note that, in a real system, the vertical field component is typically stronger for a disk that is well-coupled throughout, and, in addition, a radial field component is also generally present; e.g., \citealt{KoniglSalmeron11}). We still consider just the 2D limit. We present results for two models: a uniform disk ($a=a_s=0$) with a ``force-free'' field configuration ($q_z=0$, $q_\varphi=1$), in which the contributions of both $B_z$ and $B_\varphi$ to the radial magnetic force density vanish, and a self-similar disk ($q_z=q_\varphi=5/4$, $a=3/2$, $a_s=1/2$) akin to that considered in Section~\ref{2Dsims_bz}. For reference, we also performed simulations of a pure-$B_\varphi$ disk model, complementing the analytic results for this field configuration given in Section~\ref{basicformulation}. 
We do not discuss this case in detail since similar calculations were previously presented in \citet{Terquem03} and \citet{FromangEtal05}. However, we show relevant plots for this model in Figures~\ref{fig:tors_bz}--\ref{fig:2DFFTorq}.

\subsubsection{Force-Free Field in a Uniform Disk}
\label{forcefreesection}
\begin{figure}[ht]
\includegraphics[width=0.5\textwidth]{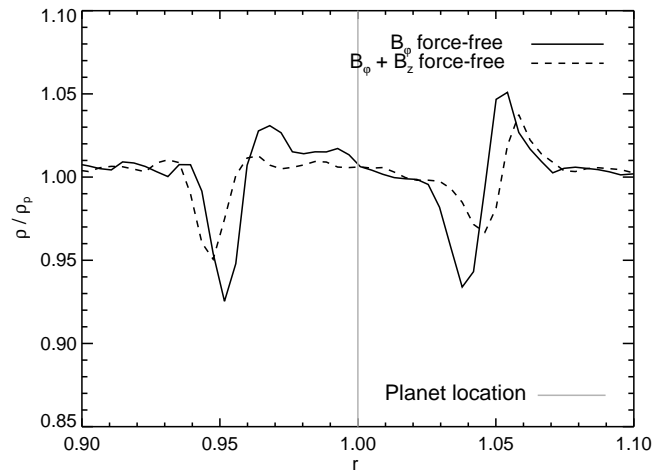}
\caption{Midplane density structure in the vicinity of the planet for two strict-2D field configurations. The dips in the plots mark the locations of the magnetic resonances. The parameters of the uniform, force-free disk model are $a=0$, $a_{s}=0$, $q_{z}=0$, $q_{\varphi}=1$, and each magnetic field component is characterized by $\beta_{\rm p}=1$.}
\label{fig:plot1}
\end{figure}

As we pointed out in Section~\ref{pure2Dsec}, the numerical implementation of the 2D limit is in general more constraining than the linear-analysis condition $k_z\rightarrow 0$, and, for the field configuration we now consider, can be mimicked only by setting $v_z^\prime=0$ in the analytic derivation. Our discussion in Section~\ref{pure2Dsec} revealed that the presence of an azimuthal field component in a 2D disk introduces magnetic resonances, and that the addition of a vertical field component shifts their locations away from $r_{\rm c}$ (which, for the parameters adopted in this subsection, is equal to $r_{\rm p}$; see Equation \eqref{eq:rcorot2}). The last 2 panels in Figure~\ref{fig:rphi_plots_rho} demonstrate the appearance of the MRs when a $B_\varphi$ component is present, and Figure~\ref{fig:plot1} confirms the predicted outward shift in the resonance positions when a $B_z$ component is added. In fact, according to Equation \eqref{2Dres}, for the chosen parameter values $r_{\rm MR} = r_{\rm p} \pm 0.047$ for the pure-$B_\varphi$ case and $r_{\rm MR} = r_{\rm p} \pm 0.06 6$ for the $B_z+B_\varphi$ configuration. These compare well with the locations of the resonant features in the density plots shown in Figure~\ref{fig:plot1}. The radial profile of the resonant features in each of these plots follows the structure of a tightly wound spiral perturbation that does not propagate radially, and so it changes with azimuth; in the example shown in Figure~\ref{fig:plot1}, the curves correspond to an azimuthal cut that is close to the planet's location ($\phi_{\rm p}-\phi_{\rm cut}=0.09$).

\begin{figure*}[ht]
 \begin{minipage}[b]{0.45\linewidth}
\includegraphics[width=\textwidth]{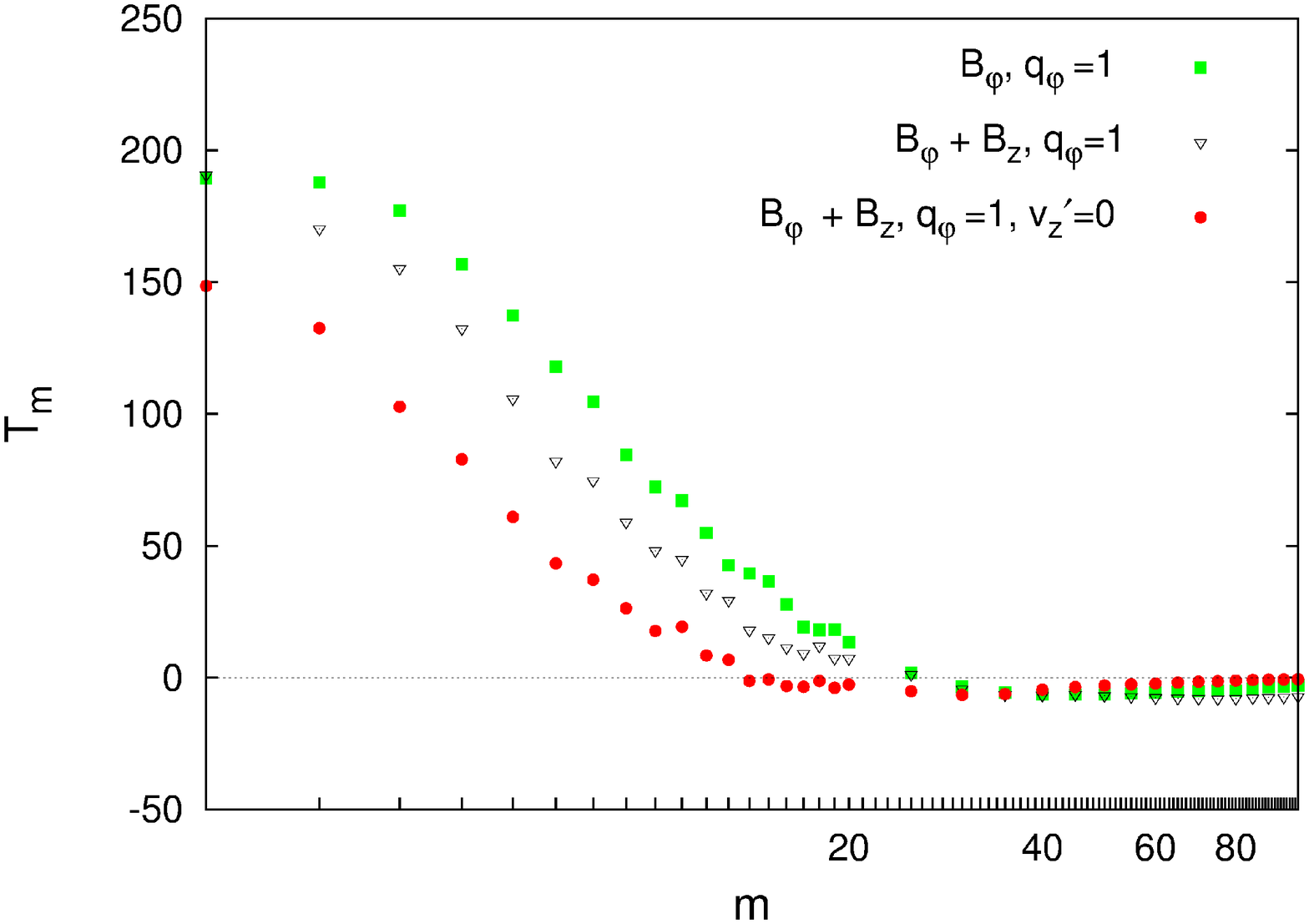}
\end{minipage}
\begin{minipage}[b]{0.49\linewidth}
\includegraphics[width=\textwidth]{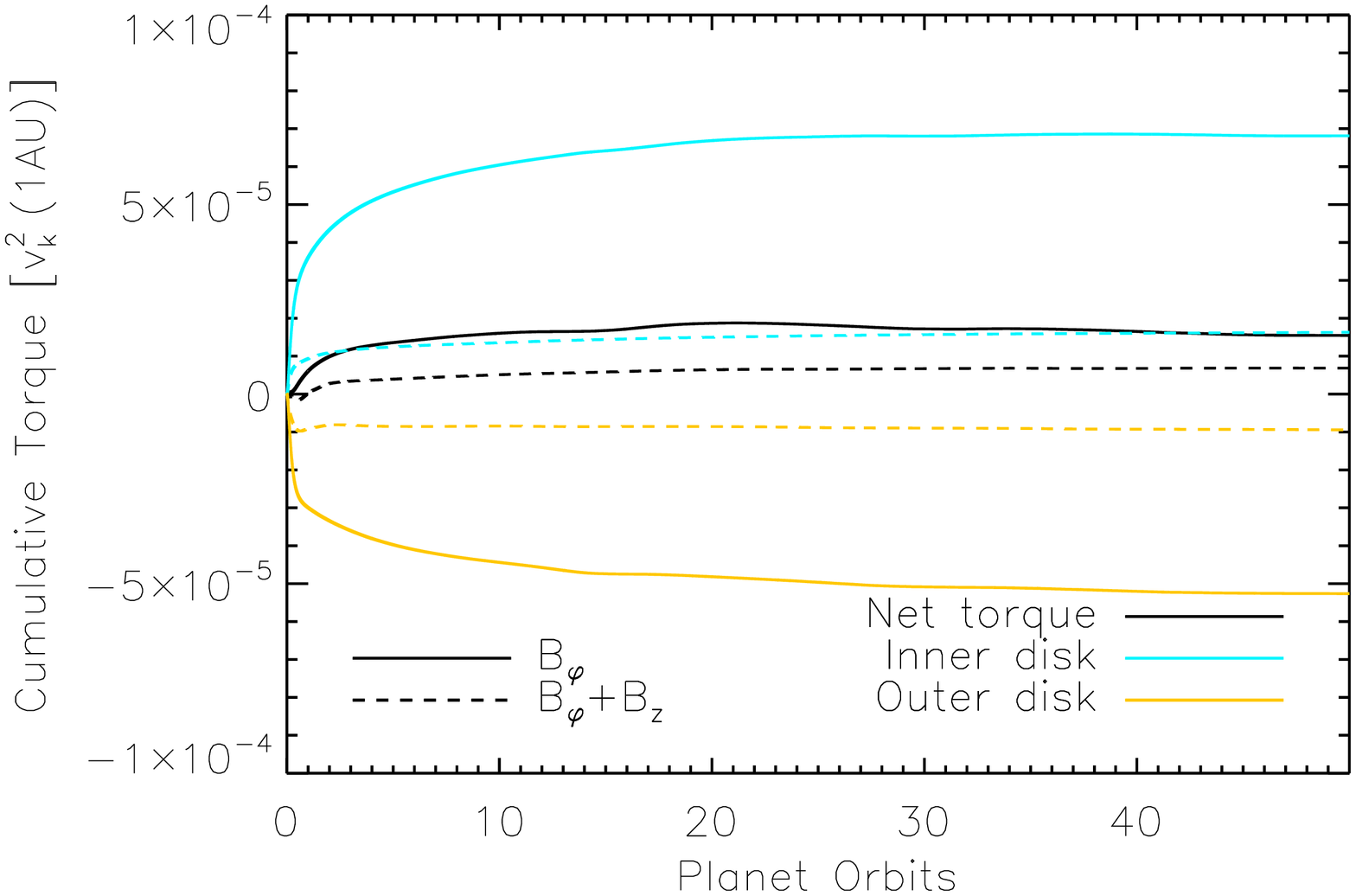}
\end{minipage}
\caption{Left: Torque on the planet  as a function of the azimuthal mode number $m$, calculated  semianalytically for a 2D pure-$B_{\varphi}$ field and for a $B_{\varphi} +  B_{z}$ field configuration assuming either $k_z=0$ (but with $v_{z}^{\prime}$ allowed to be nonzero; 2D) or  $v_{z}^{\prime} = 0$ (strict 2D). Right: Cumulative torque on the planet from the 2D numerical simulations for the same two field configurations shown in the left panel. Both calculations employ the same radial scalings ($a=0$, $a_{s}=0$, $q_{z}=0$, $q_{\varphi}=1$) and assume $\beta_{\rm p}=1$ for each field component, although they differ in the value of $c_{\rm p}$ (0.03 in the semianalytic treatment vs. 0.1 in the simulations). The torque units used in the left panel can be converted into the units employed in the right panel through a multiplication by $(\Sigma_{\rm p}r_{\rm p}^{2})/M_{*}\approx (2\rho_{\rm p}h_{\rm p}r_{\rm p}^2)/M_{*}$.}
\label{fig:2DFFTorq}
\end{figure*}

Figure~\ref{fig:2DFFTorq} shows the cumulative torque exerted on the planet, calculated semianalytically (left panel) after obtaining the perturbed density from an integration of Equation \eqref{Diff} (see Section~II.3 for details), and numerically (right panel) from the simulation results.
In the left panel, the torque is plotted as a function of the azimuthal mode number $m$, whereas in the right panel it is shown as a function of time. Both derivations show that the net torque for either the pure-$B_\varphi$ or the $B_z+B_\varphi$ field configuration is positive, corresponding to {\emph{outward} migration. The result for the pure-$B_\varphi$ with $q_\varphi=1$ field in a uniform disk was previously obtained by \citet{Terquem03} (see top panel of figure 6 in that paper), and the fact that the same trend is exhibited by the $B_\varphi+B_z$ field configuration is not surprising in view of the turning-point analysis in Section \ref{TPs}, which indicated that the addition of a vertical field component weakens the torque but does not change the basic behavior of the system. The results presented in Figure~\ref{fig:2DFFTorq} quantify the extent to which the addition of the $B_z$ component reduces the net torque. The mitigating effect of an added vertical field component is also evident in the weakening of the density perturbation at the MR locations in the $B_\varphi + B_z$ model in comparison with the pure-$B_\varphi$ case in Figures~\ref{fig:rphi_plots_rho} and~\ref{fig:plot1}.

In the left panel of Figure~\ref{fig:2DFFTorq} we also present the torque for a $B_z+B_\varphi$ field configuration in the case where $k_z\rightarrow 0$ but the condition $v_z^\prime=0$ is not enforced. As predicted by the analysis in Section~\ref{TPs}, the net torque is somewhat larger in this case (implying faster outward migration) than in the strict-2D limit.

\subsubsection{Self-Similar Disk}
\label{selfsimilarsection}

\begin{figure*}[ht]
\begin{minipage}[b]{0.49\linewidth}
\centering
\includegraphics[width=\textwidth]{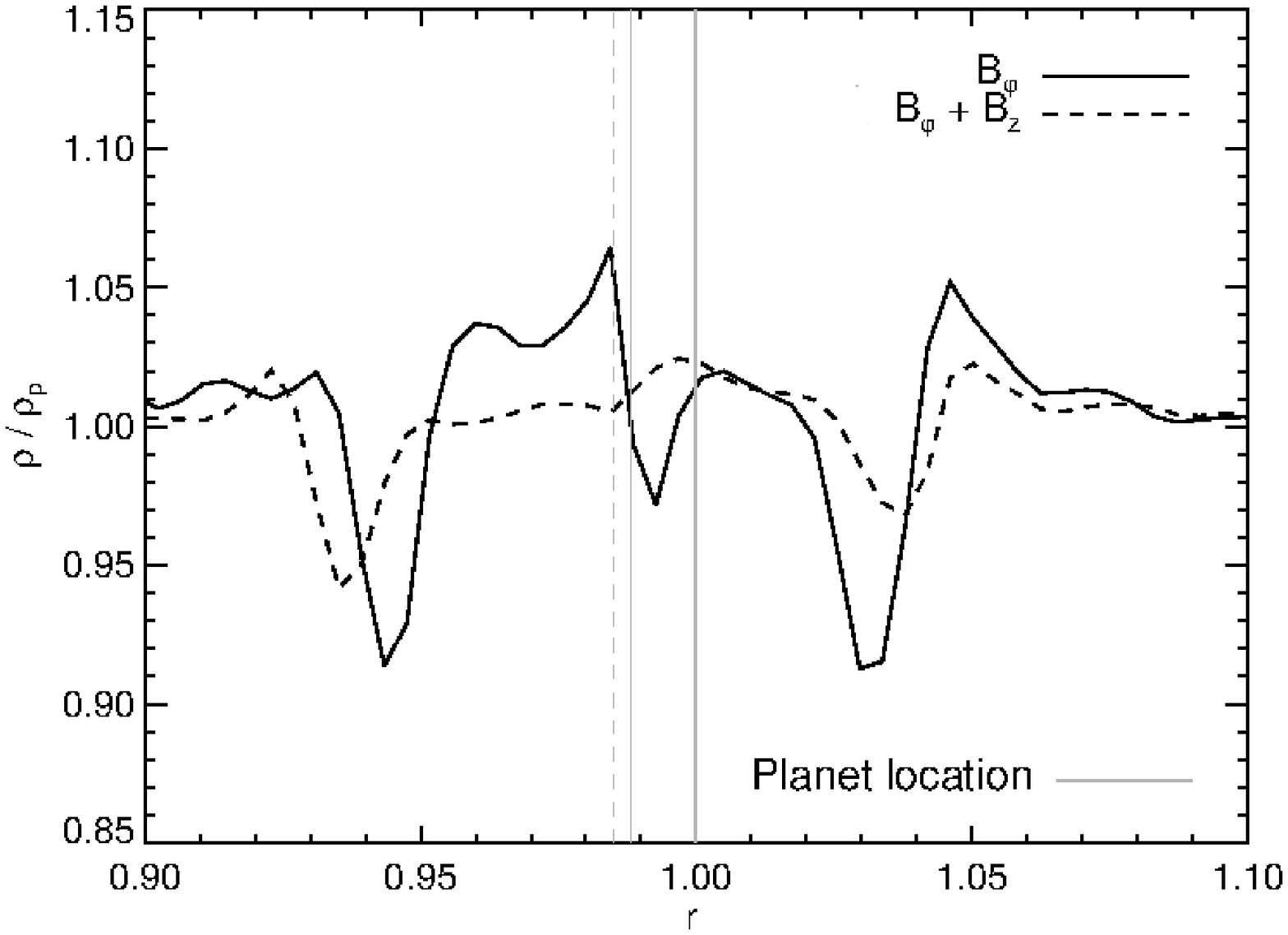}
\end{minipage}
\begin{minipage}[b]{0.49\linewidth}
\centering
\includegraphics[width=\textwidth]{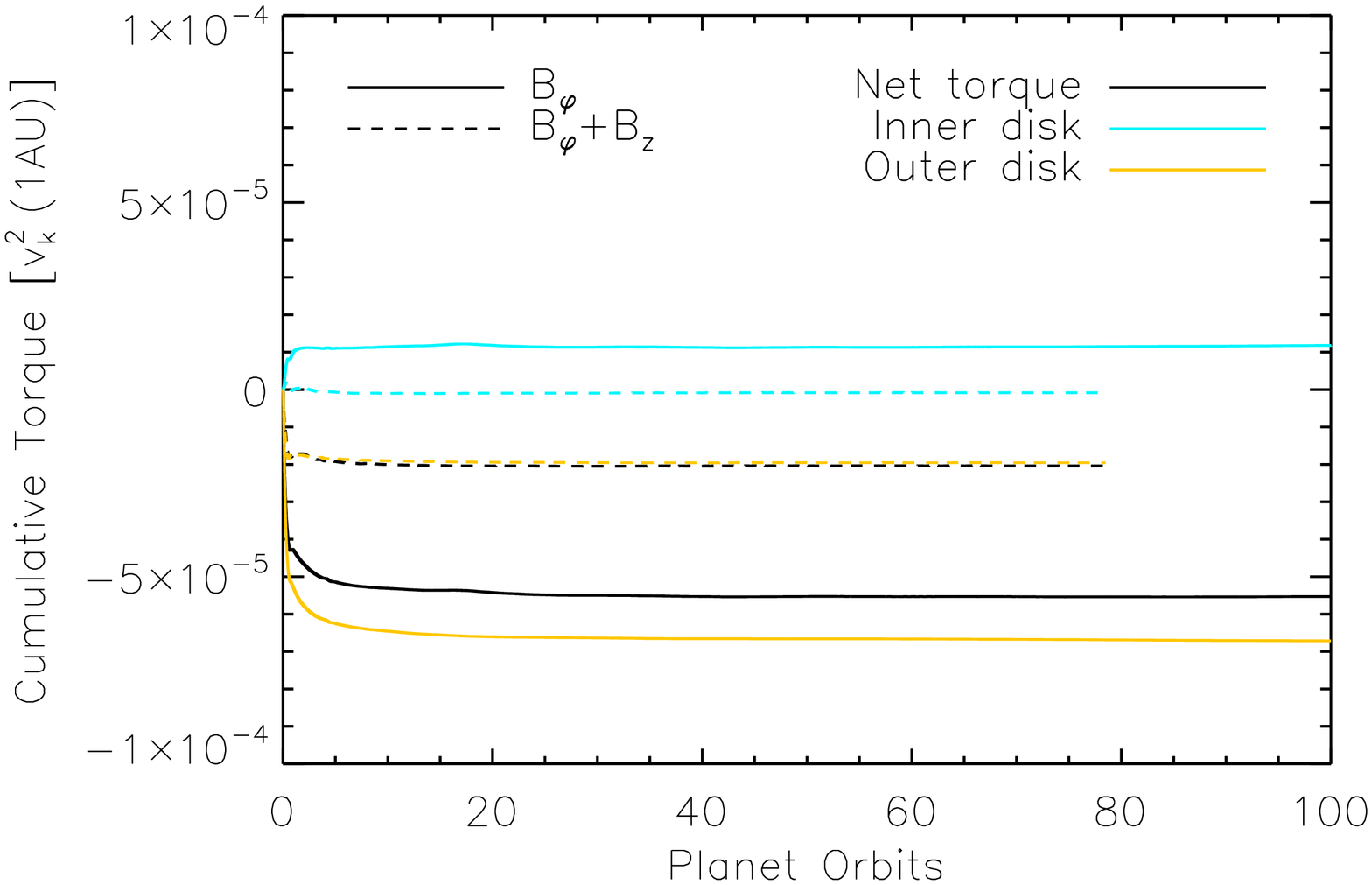}
\end{minipage}
\caption{Left: Same as Figure~\ref{fig:plot1}, except that the model parameters in this case correspond to a self-similar disk ($a=3/2$, $a_{s}=1/2$, $q_{z}=q_{\varphi}=5/4$). The locations of the corotation radii for the pure-$B_\varphi$ and $B_\varphi +B_z$ field configurations are also marked (cf. Figure~\ref{fig:rcorot}). Right: Same as the right panel of Figure~\ref{fig:2DFFTorq}, but for a self-similar disk model.}
\label{fig:plot2}
\end{figure*}

We carried out similar calculations to those in Section~\ref{forcefreesection} for this case. Our results are presented in Figure~\ref{fig:plot2}, in which the left and right panels are the analogs of Figure~\ref{fig:plot1} and the right panel of Figure~\ref{fig:2DFFTorq}, respectively. The semianalytic calculations shown in the left panel of Figure~\ref{fig:plot2} demonstrate that, in this case, too, the magnetic resonances associated with the presence of a $B_\varphi$ component are shifted away from the corotation radius when a $B_z$ component is added. However, in contrast with the situation in a uniform, force-free disk, in this case $r_{\rm c}$ does not coincide with $r_{\rm p}$ (see Equation \eqref{eq:rcorot2}  and Figure~\ref{fig:rcorot}), and it is evident from the figure that the resonances are centered approximately on $r_{\rm c}$ rather than on $r_{\rm p}$. 

The density profile shown in the left panel of Figure~\ref{fig:plot2} exhibits a noticeable dip at corotation for the pure-$B_\varphi$ disk model. This feature (which is less than $\sim 1/2$ the strength of the dips seen at the locations of the MRs for this field configuration) could arise from a nonlinear effect since there is no linear corotation resonance for a pure-$B_\varphi$ field \citep{Terquem03}. Due to its relative weakness, this feature does not significantly affect the net torque in this example. However, the questions of its origin (including why it does not appear in the force-free case shown in Figure~\ref{fig:plot1}) and of its potential influence in other disk models merit further study.

The right panel of Figure~\ref{fig:plot2} shows that, unlike the situation considered in Section~\ref{forcefreesection}, the net torque in the self-similar case is negative, corresponding to \emph{inward} migration.
The difference from the behavior of a uniform, force-free disk can be understood as follows. We recall, first, that in a Keplerian HD disk several factors combine to make the net torque on the planet negative \citep[e.g.,][]{Ward97}, one being the fact that the outer LRs lie slightly closer to the planet than the inner resonances of the same order. In such a disk, a radial density gradient has little effect on the torque since the enhanced weighting of  a denser inner disk is compensated to a large extent by the associated thermal pressure gradient, which shifts $r_{\rm c}$ inward relative to $r_{\rm p}$ and thereby increases the relative strength of the outer resonances. In the case of a uniform, force-free disk, $\beta_\varphi\propto r^2$, resulting in the outer MR being shifted (by magnetic pressure effects) farther away from $r_{\rm c}$ than the inner MR (see Equation \eqref{eq:r_MR}), thereby enhancing the relative contribution of the inner MR. (There is no shift in $r_{ \rm c}$ with respect to $r_{\rm p}$ in this case because of the absence of thermal and magnetic pressure forces.) This has led to the outward migration that we found in Section~\ref{forcefreesection}. However, in the present case there is no variation in the positions of the inner and outer MRs relative to $r_{\rm c}$ because $\beta_\varphi$ is a spatial constant in a self-similar disk. At the same time, $r_{\rm c}$ is shifted inward relative to $r_{\rm p}$ (as the left panel of Figure~\ref{fig:plot2} explicitly demonstrates), enhancing the relative contribution of the outer MR. As a result, the net torque is negative in this case (although its magnitude is reduced by $\sim 35\%$ when a vertical field component of equal amplitude is added to the azimuthal field component). This interpretation is also useful for understanding the results presented by \citet{FromangEtal05}. In fact, we find that among the different magnetized disk models that they explore (listed in their table~1), the ones that exhibit outward migration are those in which $\beta_\varphi$ \emph{increases} with $r$. (For their remaining models, which exhibit inward migration, $\beta_\varphi$ decreases with $r$.) \citet{FromangEtal05} discuss the sense of migration in their model disks in relation to the radial variation of $B_\varphi$, but the radial dependence of $\beta_\varphi$ is a more pertinent quantity in this regard.

\subsection{3D Simulations}\label{3Dsims}

We now present results from 3D simulations that take into account the vertical component of the planet's gravity (although, unless stated otherwise, we continue to neglect the vertical component of the stellar gravitational field). We explore the vertical structure of resonances and the effects of wave propagation in the $z$ direction on the direction and speed of planet migration. We note that our 3D simulations do not have the numerical resolution to separate the MRs from the ARs. However, our semianalytic study indicates that the dominant contribution to the density perturbations in this regime comes from the MRs.

\subsubsection{Vertical Field in a Uniform Disk}\label{3dsims_bz}

This case is studied semianalytically in Paper~II. Figure \ref{fig:rho_bz_3d} presents the midplane density structure from a numerical simulation performed using as model parameters $a=0$, $a_{s}=0$, $q_{z}=0$, and $\beta_{z\rm p}=1$. The plot exhibits a dominant one-armed spiral (with an amplitude of  $\sim 10\%$ of the unperturbed disk density)
as well as similar spiral features that are, however, much weaker
($< 1\%$ of $\rho_0$). The stronger spiral resembles the perturbed density structure of a Keplerian HD disk and can be similarly interpreted as a wake formed by the constructive interference of acoustic waves of different azimuthal mode numbers $m$ that are launched at the effective Lindblad resonances \citep[e.g.][]{OgilvieLubow02}. We interpret the fainter spiral features as analogous ``magnetic wakes'' that are formed from the constructive interference of SMS and Alfv\'en waves launched at the turning points associated with the MRs and ARs. In fact, according to Equations \eqref{ARFull} and \eqref{MRFull} (see also Figure~II.1), both ARs and MRs are present in this case, and their locations depend on $m$. The locations of the associated turning points are likewise $m$-dependent (Figure~II.2). 

As shown by the short-dashes black curve in Figure~\ref{fig:cum_tors_all}, the net torque in this case is negative --- corresponding to inward migration --- but is reduced by $\sim 50\%$ with respect to the 3D torque exerted on the planet in a uniform, unmagnetized disk (solid black curve). This result is qualitatively similar to our finding for this field configuration in the 2D limit (Section~\ref{2Dsims_bz}, although note that we considered a different disk model in that case).
One can understand this similarity on the basis of our semianalytic results (Section~II.4.2), which indicate that the cumulative 3D torque is comparable to the integrated contribution of the $k_z=0$ vertical mode (i.e., to the 2D torque) for the pure-$B_z$, uniform disk model. 
In Paper~II we obtain a rough semianalytic estimate for the net torque  in this case --- a factor of $\sim 2$ smaller than the 2D HD torque.
We have also estimated semianalytically the net torque for an unmagnetized disk in 3D and found that is slightly smaller than the 2D HD 
torque.\footnote{\citet{TanakaEtal02} found that, in the HD case, the 2D torque exceeds the 3D torque by a factor of $\gtrsim 2$. However, they employed different disk parameters, and therefore an exact correspondence with our result is not expected.} Thus, in both 2D and 3D, we estimate that the net torque on disks with a pure-$B_{z}$ field is a factor of $\sim 2$ smaller than the HD torque. This is consistent with our 2D and 3D simulation results.
\begin{figure}[ht!]
\begin{minipage}[b]{0.5\textwidth}
 \centering
   \includegraphics[width=0.6\textwidth]{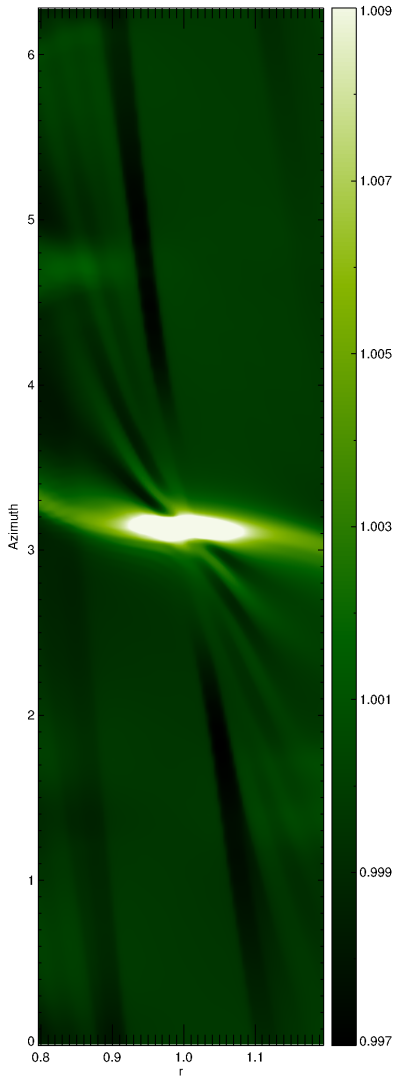}
\end{minipage}
   \caption{Midplane density in a uniform, 3D disk permeated by  pure-$B_z$ field with $\beta_{z\rm p}=1$. The planet is located at $r=1.0$ and $\varphi=\pi$. The contrast in the figure was adjusted to show, in addition to the bright Lindblad wake formed by the constructive interference of acoustic (FMS) waves, also the analogous, but much fainter, wakes that are formed by SMS and Alfv\'en waves.}
\label{fig:rho_bz_3d}
\end{figure}

\begin{figure*}[ht!]
\centering
   \includegraphics[width=0.8\textwidth]{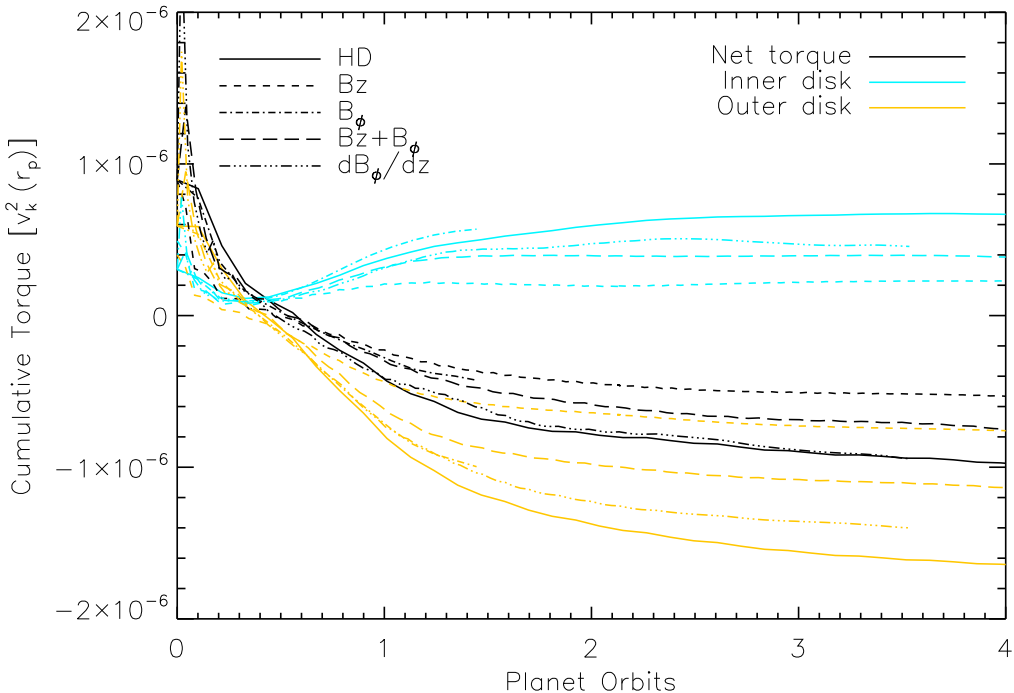}
\caption{Cumulative torque for all of the reported 3D simulations of an unstratified disk.}\label{fig:cum_tors_all}
\end{figure*}

\subsubsection{Force-Free Azimuthal Field in a Uniform Disk}\label{3dsims_bphi}

This model ($a=0$, $a_{s}=0$, $q_{\varphi}=1$) extends the 2D treatment of a pure-$B_\varphi$ disk in \citet{Terquem03} and \citet{FromangEtal05} to 3D. It incorporates the perturbing vertical gravitational force exerted by the planet and the magnetic and thermal-pressure forces arising from the disk's response.

The behavior of the midplane density for this case is shown in Figure~\ref{fig:plot_grid_rho}. During the first few orbital periods of the planet, the density structure appears to evolve similarly to that in the 2D simulation (Figure~\ref{fig:rphi_plots_rho}). However, in contrast to the 2D case, the net torque becomes negative (dash-dotted black curve in Figure~\ref{fig:cum_tors_all}). The reason for this difference is probably similar to the cause of the corresponding behavior of the $B_\varphi+B_z$ field configuration, which we discuss in Section~\ref{3dsims_bzbphi}.
In any case, after about 3 to 4 orbits, even before the imprints of the resonances have been fully established, the disk develops an instability roughly at the radial position of the the planet. This instability spreads rapidly to the rest of the disk (mostly into the inner regions over the duration of the simulation), creating turbulence that becomes the dominant influence on the subsequent evolution of the planet.
This evolution is not pursued in this paper as it would require a significant additional computational effort \citep[e.g.,][]{UribeEtal11}. For this reason we have truncated the hybrid-field curve in Figure~\ref{fig:cum_tors_all} just before the onset of the instability.

A more detailed characterization of the observed instability is as follows. 
Initially, the planet creates small perturbations around its orbit in all the physical quantities (density, velocity, and magnetic field). The vertical field component starts growing after about 1.5 orbits, and the radial field component follows suit after $\sim 2.5$ orbits (see left panel of Figure~\ref{fig:magene_evol}). The two poloidal field components then grow exponentially together, increasing in energy by $\sim 4-5$ orders of magnitude until they saturate after $\sim 6$ orbits. The evolution of the associated $\beta$ parameters is shown in the right panel of  Figure~\ref{fig:magene_evol}.
The figure also shows that the energy associated with the azimuthal field, which is dominated by the ordered (equilibrium) component of $B_\varphi$, starts decreasing when the instability is triggered, with the decline becoming more pronounced after the instability saturates. We have verified that this decline is associated with the radial mass inflow that is induced by the onset of MRI turbulence: the large-scale magnetic field is then simply advected inward by the inflowing mass and leaves the computational domain through the inner radial boundary. We did not attempt to address this issue in our simulations because, on the one hand, it does not affect our conclusions about the development of the instability (whose growth occurs on the dynamical timescale, which is much shorter than the radial inflow time), and, on the other hand, we do not attempt to model the longer-term evolution of the disk accurately. 

The evolution of the resulting turbulent stresses in the disk is shown in Figure~\ref{fig:alpha_sim1}: They start growing exponentially after $\sim 4$ orbits, and saturate after $\sim 6$ orbits. It is seen that the Maxwell and Reynolds stresses reach comparable magnitudes (with the Maxwell stress being larger overall). The total stress (normalized by the initial thermal pressure at $r_{\rm p}$ and presented as an effective viscosity parameter) grows to $\alpha \approx  0.1$.\footnote{We consider only the $r \varphi$ component of the stress and use the expressions in \citet{Hawley00} for the vertically and azimuthally averaged Reynolds and Maxewell contributions at any given radial position.}

To check whether the instability might be stabilized by vertical density stratification, we
ran a simulation with the same numerical configuration except that we also included the vertical component of the stellar gravity. We found that the instability still developed in the resulting stratified disk, and that the magnitude of the torque exerted on the planet before the development of the instability was not significantly different from the value obtained for the unstratified disk.  
The difference between these two cases is illustrated in Figure~\ref{fig:ver_struc_bphi}, which shows the poloidal ($r-z$) density structure of the MRs that develop inside and outside the planet's orbit \emph{before} the instability is triggered. The left panel of this figure demonstrates that the radial position of each of the MRs is constant with height for the uniform disk, $z$, whereas the right panel shows the fanning-out of the resonance in the stratified disk. This behavior is a direct consequence of the dependence of $r_{\rm MR}$ (Equation \eqref{eq:r_MR}) on $\beta_\varphi$ and the fact that $\beta_\varphi \propto \rho$ decreases with $z$ in a stratified disk. The observed density structure also reflects the weakening of the perturbing gravitational potential with distance from the planet, which causes the resonance amplitude to decline with height.

We discuss the nature of this instability in Section~\ref{discuss}, where we argue that it is essentially a form of the MRI. The important implication of this result to planet migration is that the conclusions drawn for this field configuration from 2D studies 
\citep[e.g.,][Section~\ref{forcefreesection}]{Terquem03,FromangEtal05}, and in particular the prediction of a systematic outward migration for large enough values of $q_\varphi$, are not applicable to real disks, which are inherently 3D. A planet embedded in a disk that is permeated by a pure-$B_\varphi$ field would evolve, instead, according to the predictions of MRI-turbulent disk models \citep[e.g.,][]{NelsonPapaloizou04}. In particular, a planet with the mass employed in our simulations ($\sim 5 M_\earth$) would experience the torque responsible for Type I migration in a laminar disk as well as a measurable random torque component induced by the turbulence, likely resulting in a nonmonotonic, but overall inward-directed, drift. The relative impact of the stochastic component is predicted to increases with decreasing planet mass \citep[e.g.,][]{UribeEtal11}.

\begin{figure*}[ht!]
 \begin{minipage}[b]{\textwidth}
 \centering
 \includegraphics[width=\textwidth]{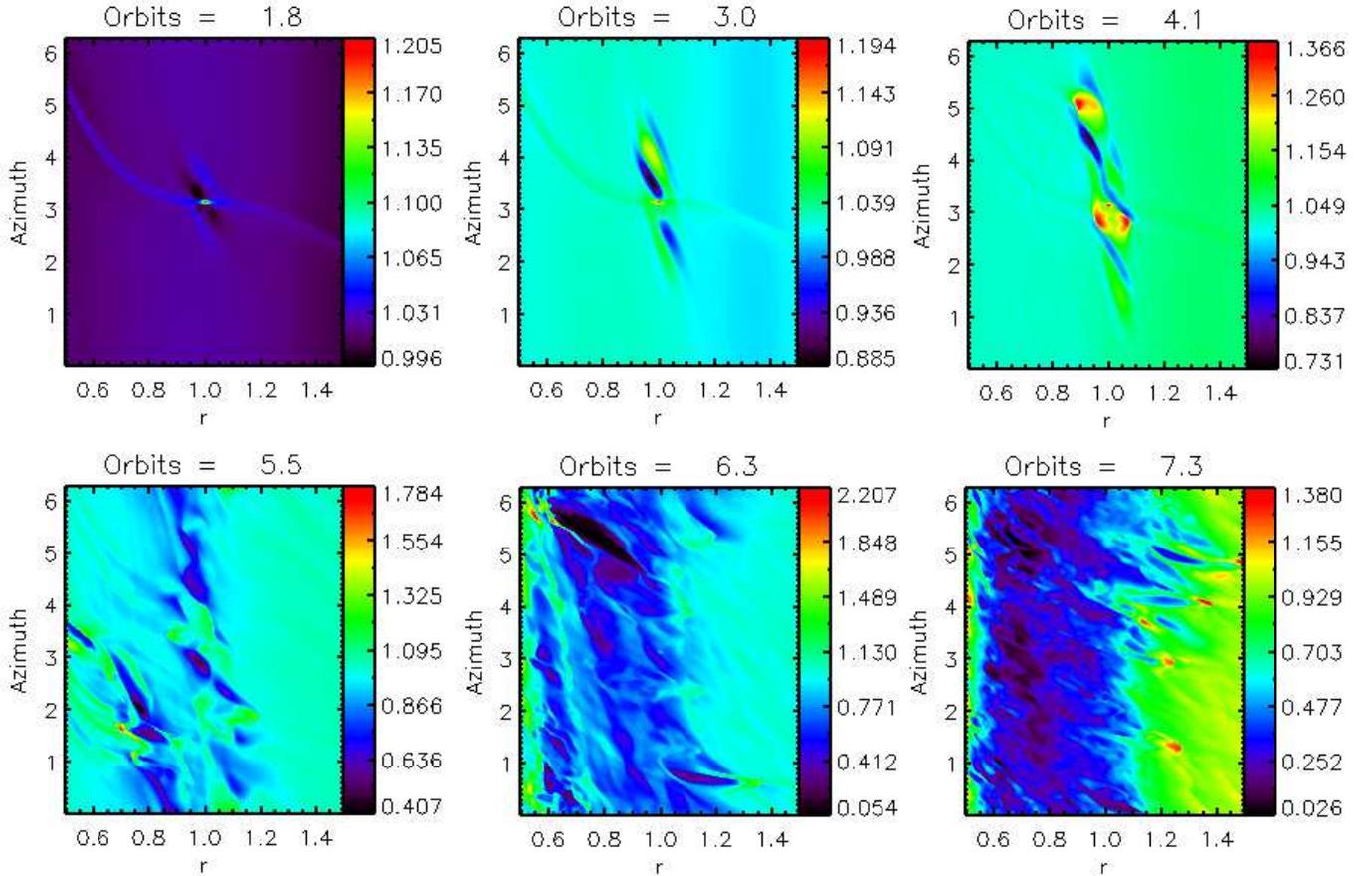}
\caption{Snapshots of the midplane density (labeled by the number of planet orbits) in a uniform, 3D disk with a force-free $B_\varphi$ field. The planet is located at $r=1.0$ and $\varphi=\pi$. The development of an instability after $\sim 3-4$ orbits is clearly visible.}
\label{fig:plot_grid_rho}
\end{minipage}
\end{figure*}

\begin{figure*}[ht]
\begin{minipage}[b]{0.49\linewidth}
\centering
\includegraphics[width=\textwidth]{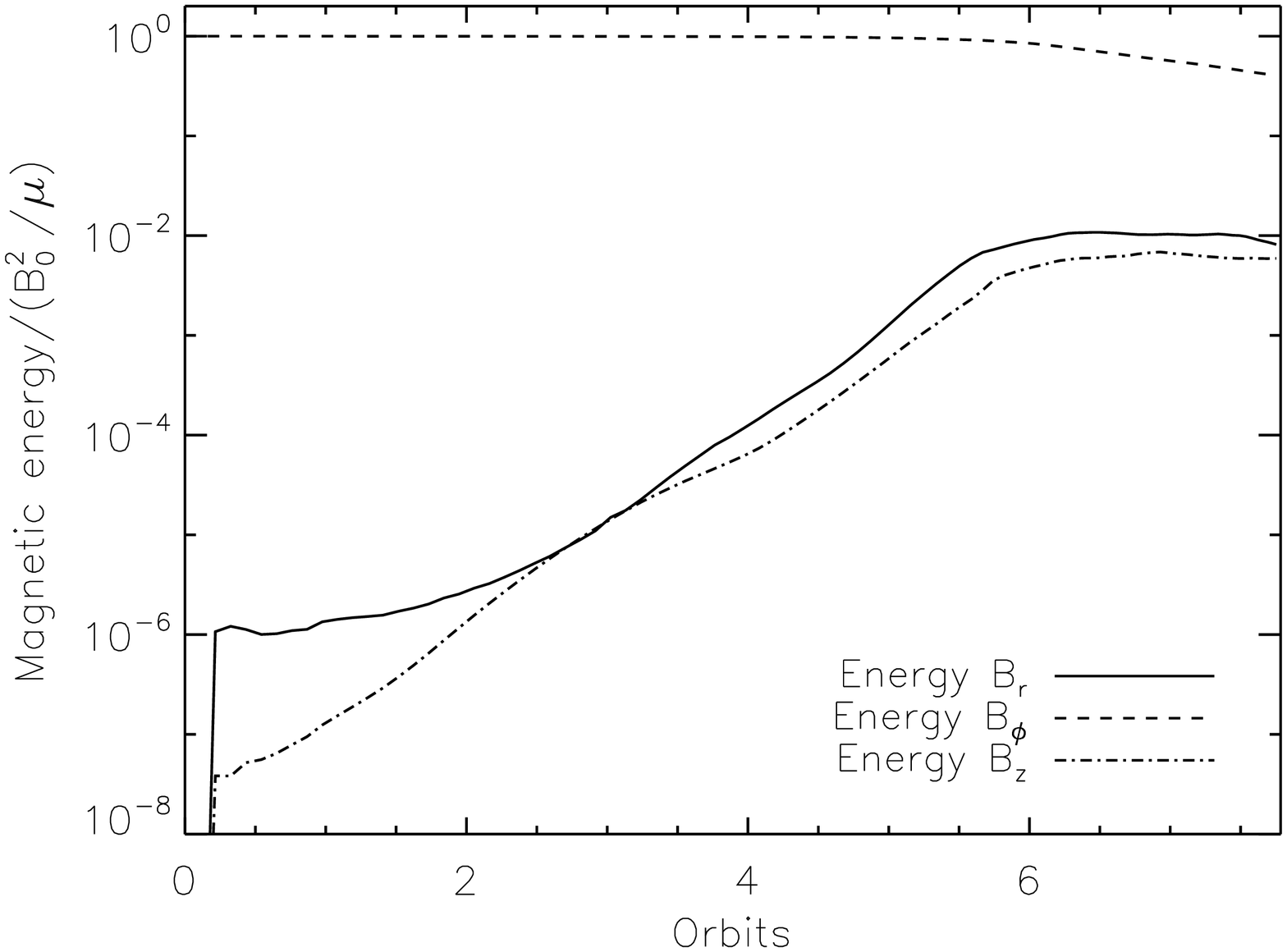}
\end{minipage}
\begin{minipage}[b]{0.49\linewidth}
\centering
\includegraphics[width=\textwidth]{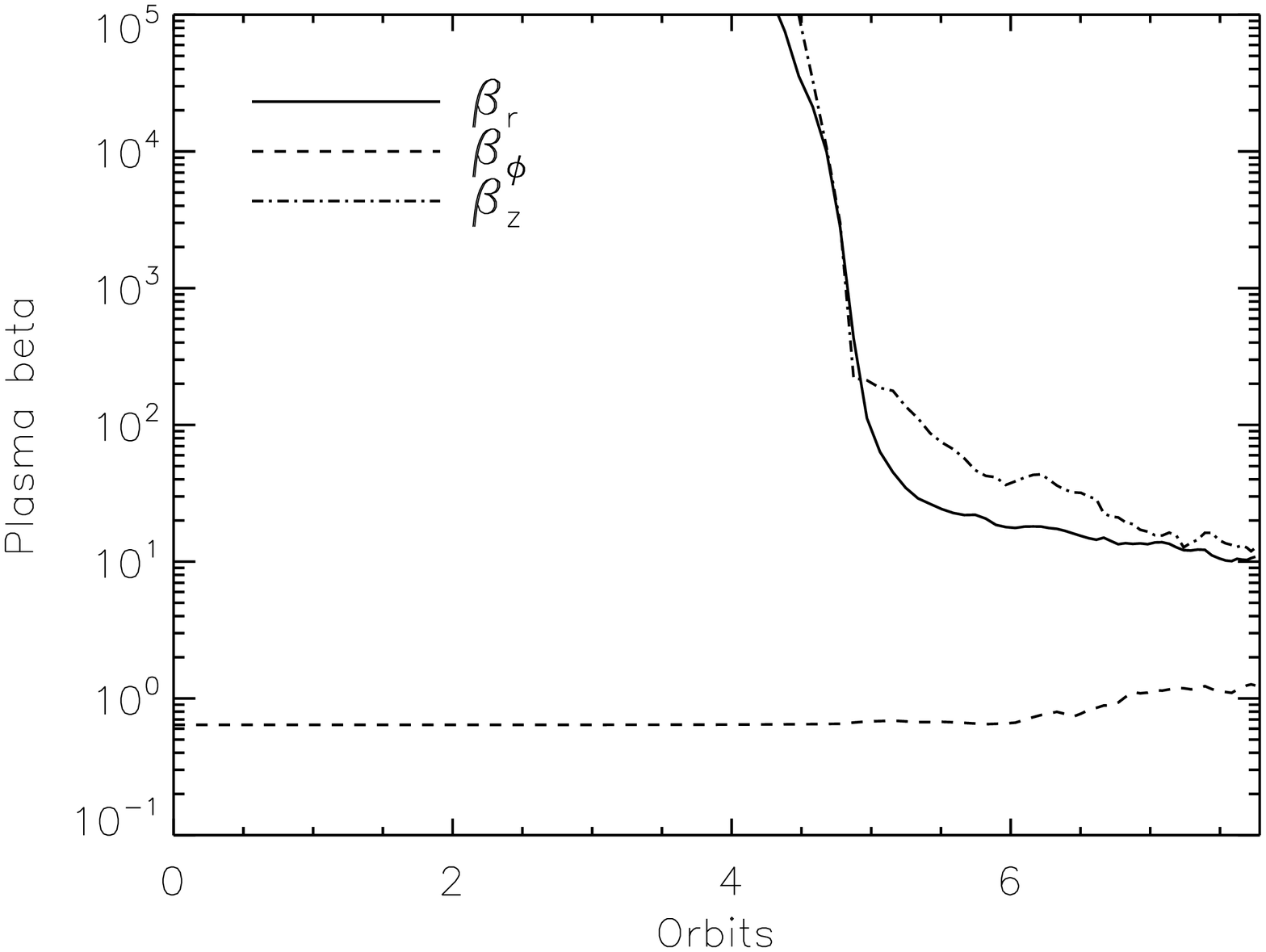}
\end{minipage}
\caption{Time evolution of the magnetic energy associated with each field component (Left) and of the corresponding $\beta$ parameters (Right), spatially averaged over the full computational domain, for a 3D, uniform disk with a force-free $B_\varphi$ field.}
\label{fig:magene_evol}
\end{figure*}

\subsubsection{Disk with $B_\varphi \propto z$}\label{3dsims_bphi_zgrad}

We now consider the same disk model as in Section~\ref{3dsims_bphi} except that, instead of having a force-free azimuthal field  that is uniform in $z$ ($B_{0 \varphi}=B_{\varphi \rm p} r^{-1}$), we add a dependence on the vertical coordinate: $B_{0 \varphi}=B_{\varphi \rm p} (z/z_0)r^{-1}$, where $z_0=0.13$ is the upper domain boundary (see Section~\ref{num_setup}). The value of $B_{\varphi \rm p}$ is the same as that used in Section~\ref{3dsims_bphi} and corresponds to $\beta_{\varphi \rm p} =1$ if one uses the midplane density in evaluating this parameter. (The actual value of $\beta_\varphi$ at $z=z_0$ is, however, $<1$ since the density there is lower than at the midplane.)
The adopted vertical structure of $B_\varphi$ is intended to approximate the azimuthal field in a wind-driving disk that is threaded by open field lines \citep[e.g.,][]{WardleKonigl93}, and we consider it here (as well as semianalytically in Section~I.4.3) in order to isolate the effects of the $\partial B_\varphi/\partial z$ term on the disk behavior. The main expected impact would come from the magnetic torque term in the angular momentum equation ($(B_z /\mu)\partial B_\varphi/\partial z$) --- however, this term (which gives rise to radial motion that, in turn, generates a $B_r$ component in a disk threaded by open field lines) cannot be included in an ideal-MHD simulations that aims to approach a steady state (see Section~\ref{intro}).\footnote{We are, however, able to infer the effect of the magnetic torque term on the linearized equations; see Paper~II.}

The $z$-gradient term also appears in the vertical momentum equation ($-(\partial/\partial z)[B_\varphi^2/(2\mu)]$), where it represents magnetic pressure squeezing that has the same effect as the tidal squeezing by the stellar gravity. One can therefore anticipate that the poloidal structure of the resonances in this case will be similar to that of the stratified disk model considered in Section~\ref{3dsims_bphi}.
This is confirmed by the results shown in Figure~\ref{fig:ver_struc_bphi_zgrad}, which resemble those in the right panel of Figure~\ref{fig:ver_struc_bphi}. We initialized the simulations presented in Figure~\ref{fig:ver_struc_bphi_zgrad} using the approximate vertical-equilibrium expression derived in Appendix~\ref{appB}; the density adjusts to a fully consistent steady state after a few orbits. We show results for a model in which just the magnetic squeezing effect is included as well as for a disk in which both magnetic and tidal squeezing operate. According to Equation \eqref{eq:rhoz}, the ratio of the magnetic and tidal squeezing terms is $\sim \beta_\varphi^{-1}(h_{\rm p}/z_0)^2$, which equals 0.59 for our chosen parameter values ---- indicating that the magnetic squeezing associated with the $B_\varphi$-gradient term has only a small effect. This is consistent with more realistic models of wind-driving disks \citep[e.g.,][]{KoniglSalmeron11}, which indicate that magnetic squeezing is important in well-coupled disks but is induced by the radial, rather than the azimuthal, field component. (As we noted above, we cannot include a $B_r$ field in an ideal-MHD simulation of this problem.)

The fact that the amplitude of the azimuthal field component remains small in the vicinity of the planet weakens the effect of the resonances, which only develop above the midplane in this case. Consequently, the torque on the planet in this model is effectively the same as in an unmagnetized disk (see dash--triple-dotted black curve in Figure~\ref{fig:cum_tors_all}, where an unstratified disk is considered). This result is consistent with the inference from our semianalytic study of this field configuration (with $q_\varphi=0$ instead of 1) in Paper~II. Another consequence of the small amplitude of the equilibrium magnetic field near the midplane is that the perturbed vertical magnetic field is evidently too weak (for the given level of numerical diffusivity) to trigger the MRI, so that, in contrast with the behavior of the uniform-$B_\varphi$ disk (see Sections~\ref{3dsims_bphi} and~\ref{discuss}), no instability develops in this case.

\begin{figure}[ht!]
\begin{minipage}[b]{\textwidth}
   \includegraphics[width=0.5\textwidth]{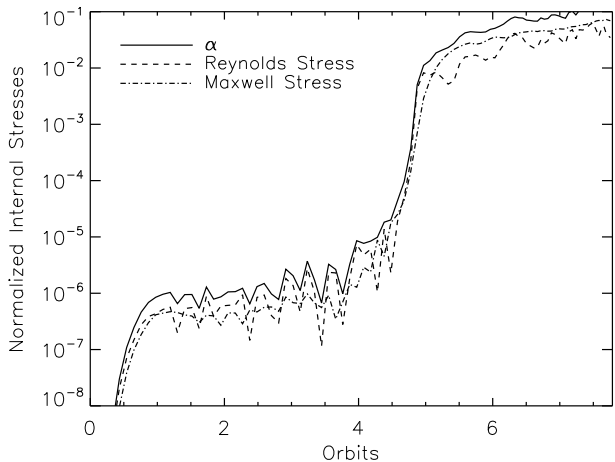}
   \end{minipage}
\caption{Time evolution of the Reynolds, Maxwell and total stresses, each normalized by the initial thermal pressure at the planet's radial distance from the star ($1\;$AU), for the model considered in Figure~\ref{fig:magene_evol}. The stresses are averaged over the azimuthal and vertical directions and are evaluated at a disk radius of $0.8\;$AU.}
\label{fig:alpha_sim1}
\end{figure}

\begin{figure*}[ht]
\begin{minipage}[b]{0.49\linewidth}
\centering
\includegraphics[width=\textwidth]{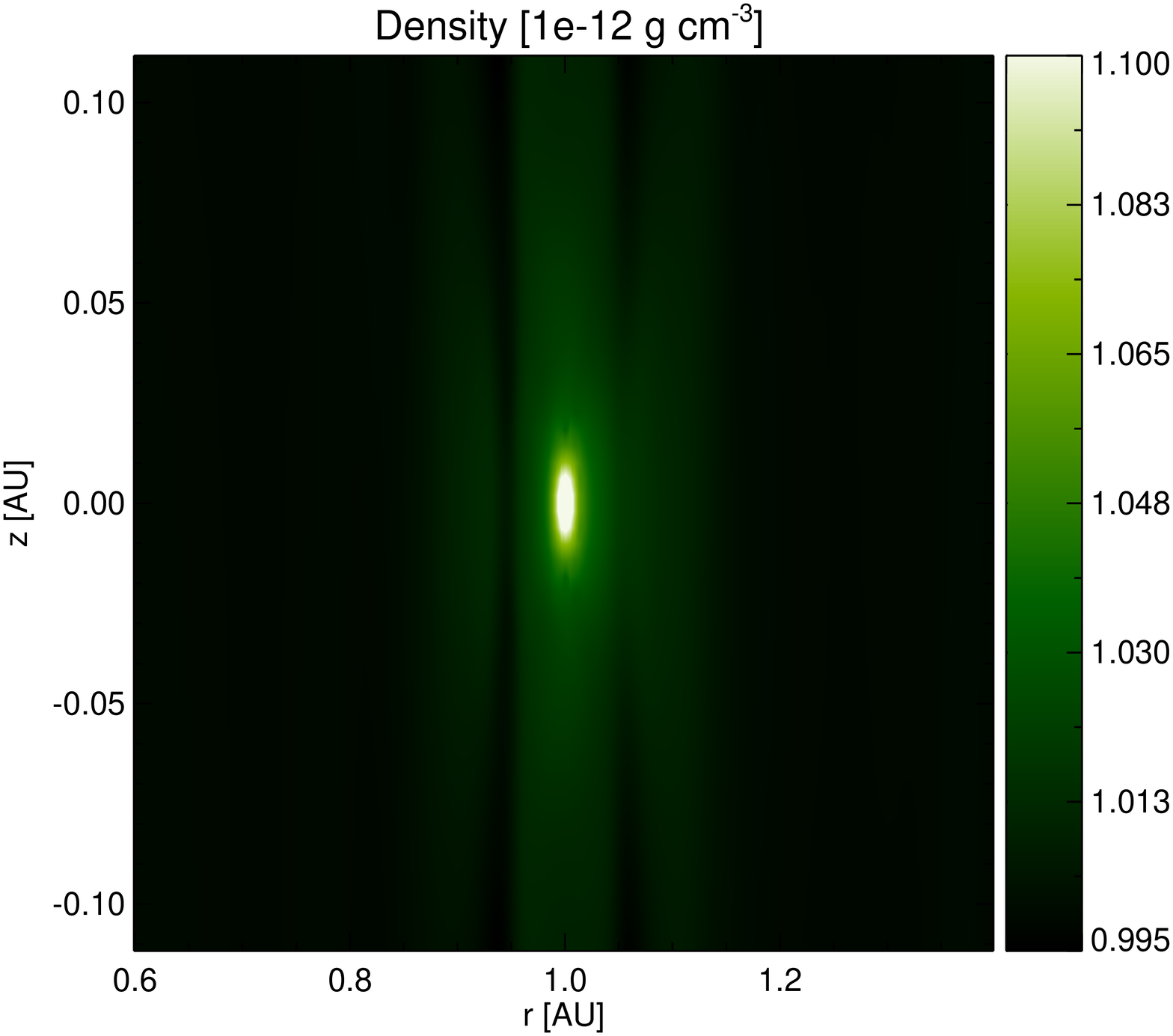}
\end{minipage}
\begin{minipage}[b]{0.49\linewidth}
\centering
\includegraphics[width=\textwidth]{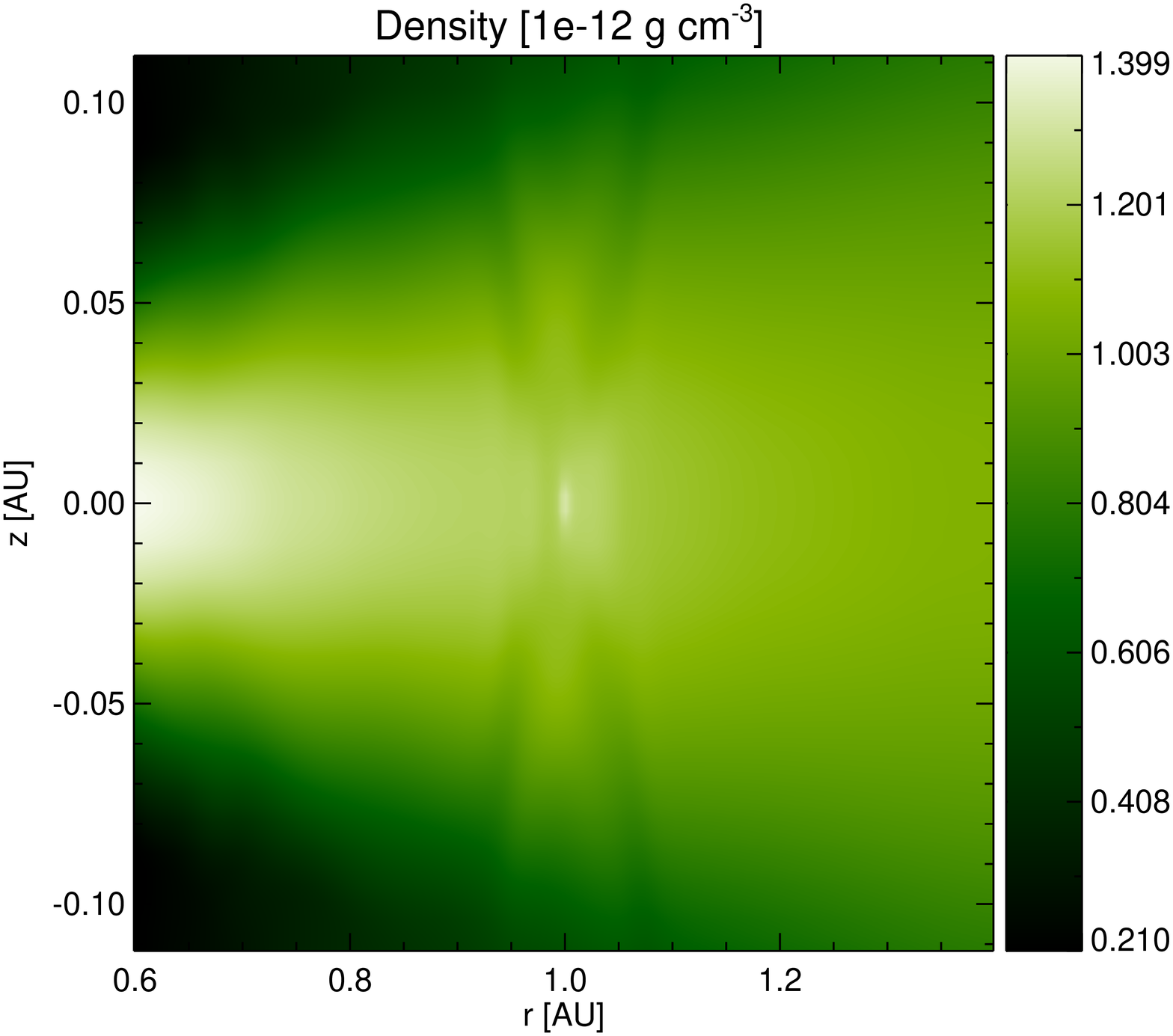}
\end{minipage}
\caption{Poloidal density structure for two 3D disk models with a force-free azimuthal field before the development of the instability shown in Figures~\ref{fig:magene_evol} and~\ref{fig:alpha_sim1}. The left panel represents a uniform disk, whereas the right panel shows a disk that is vertically stratified by the action of the stellar tidal gravity. The planet is located at $(r,z)=(1,0)$.}
\label{fig:ver_struc_bphi}
\end{figure*}

\begin{figure*}[ht]
\begin{minipage}[b]{0.49\linewidth}
\centering
\includegraphics[width=\textwidth]{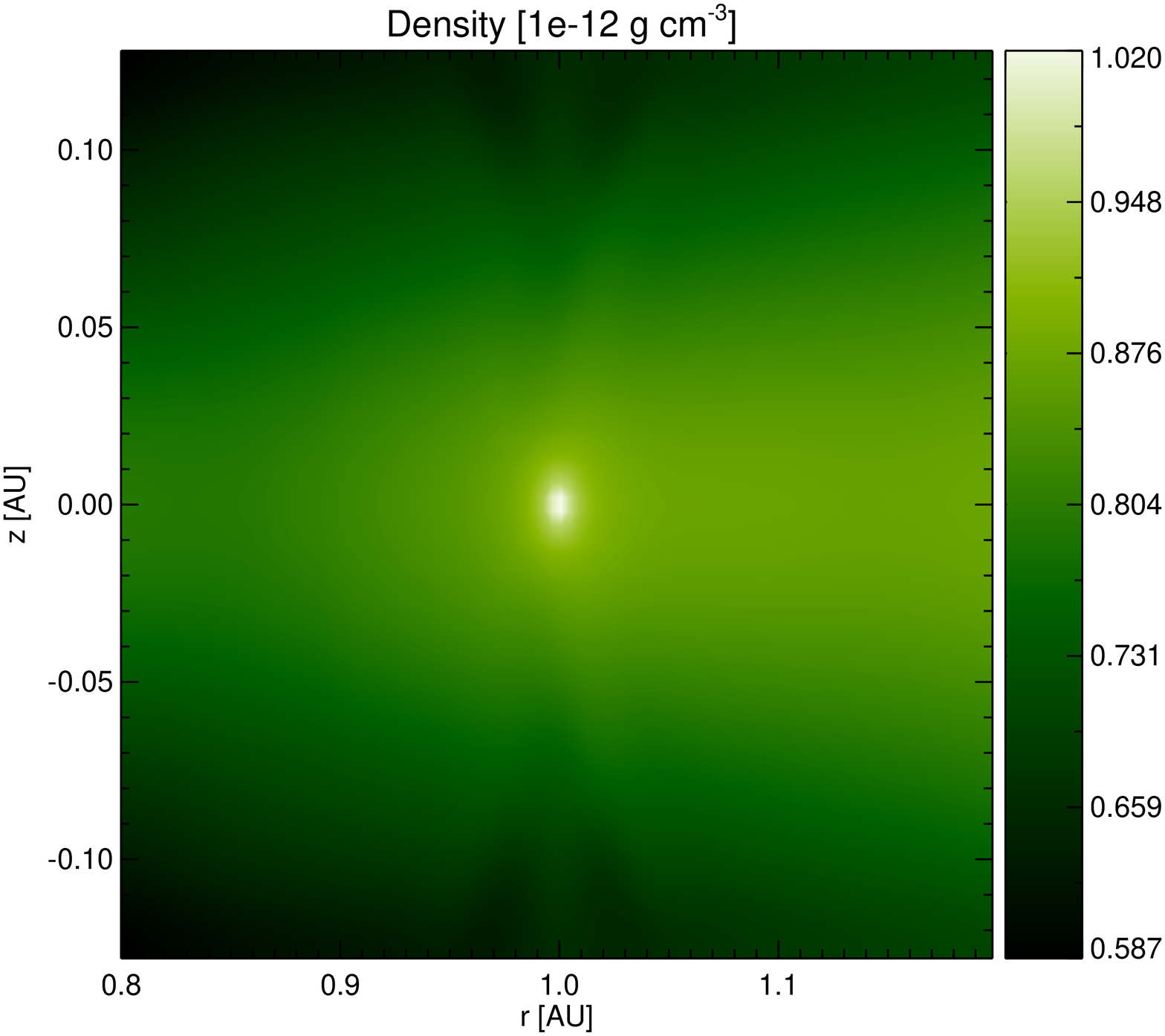}
\end{minipage}
\begin{minipage}[b]{0.49\linewidth}
\centering
\includegraphics[width=\textwidth]{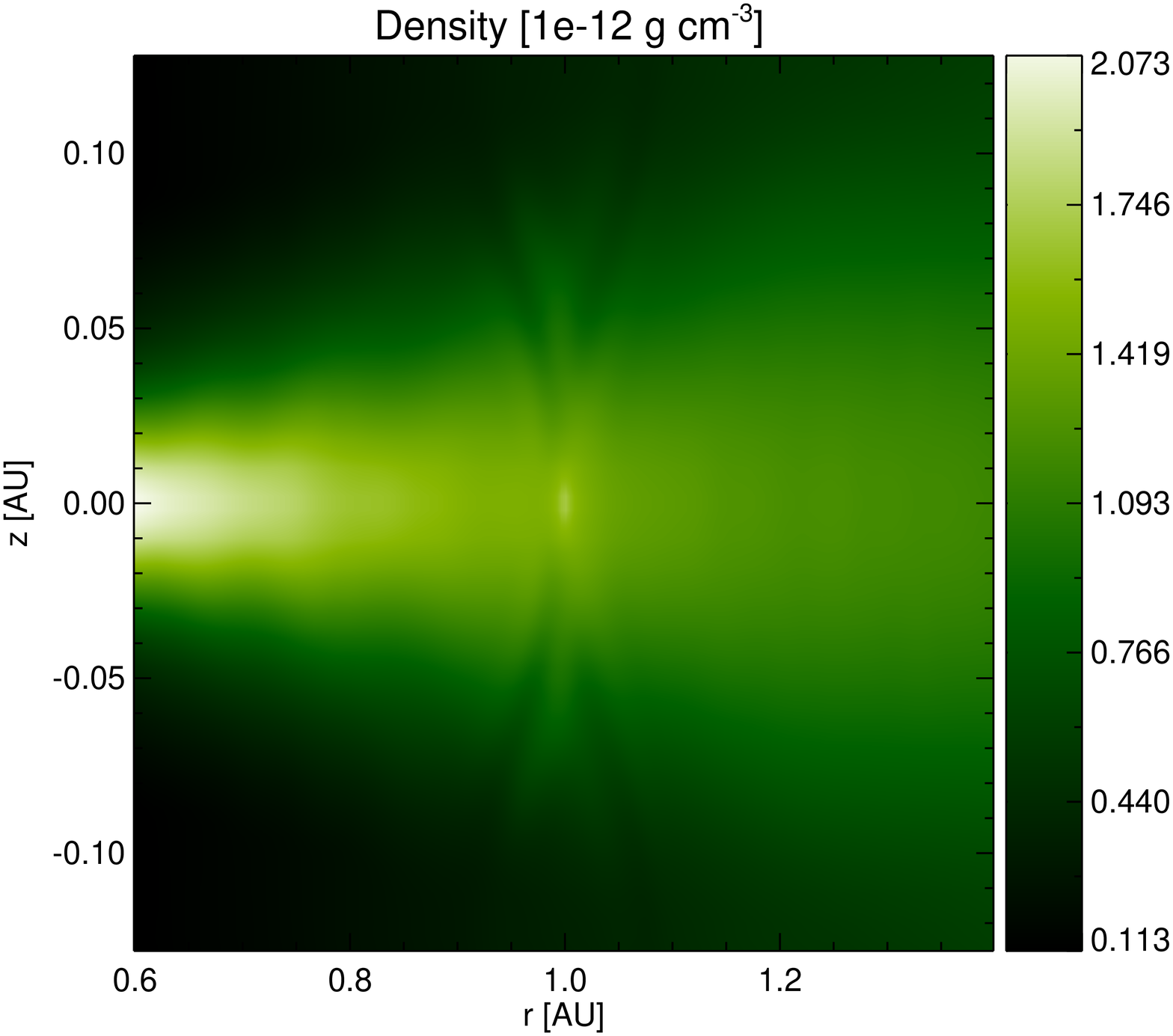}
\end{minipage}
\caption{Same as Figure~\ref{fig:ver_struc_bphi}, but for an azimuthal field that increases linearly with height from zero at the midplane to its final value (which equals its constant value in Figure~\ref{fig:ver_struc_bphi}) at the top of the computational domain. The vertical gradient of $B_\varphi$ contributes to the vertical stratification of the disk. The disk shown in the right panel includes the effects of both the field gradient and the stellar gravity and is therefore more strongly stratified than the model presented in the left panel, in which the stellar contribution is neglected. This is manifested in the larger opening angle of the MR ``perturbation cone'' in the right panel.}
\label{fig:ver_struc_bphi_zgrad}
\end{figure*}

\subsubsection{Force-Free $B_\varphi+B_z$ Field in a Uniform Disk}\label{3dsims_bzbphi}

We now combine the vertical and azimuthal field components discussed in Sections~\ref{3dsims_bz} and~\ref{3dsims_bphi}, respectively, and consider a hybrid $B_z+B_\varphi$ disk model with $a= a_{s}=0$, $q_{z}=0$, $q_{\varphi}=1$, and $\beta_{z\rm p}=\beta_{\varphi \rm p} =1$. For this field geometry, both ARs and MRs are excited --- their locations given by Equations \eqref{ARFull} and~\eqref{MRFull}, respectively --- and the associated waves (SMS and Alfv\'en) propagate along the direction of the total field. Since the resonances and turning points depend on $m$ (see the bottom panel of Figure~\ref{3DTPpanel}), the density perturbations will again form magnetic wakes, analogous to those found in the pure-$B_{z}$ case (Figure~\ref{fig:rho_bz_3d}).\footnote{In practice, the ARs are not visible or even resolved in the numerical simulations since the MR-to-AR amplitude ratio is typically $\sim 10^{4}$ for the pure-$B_{\phi}$ case and $\sim 10^{3}$ for the hybrid-field case.}

The total torque on the planet in this case is given by the black long-dashes line in Figure~\ref{fig:cum_tors_all}. Just as in the case of each of the separate components (with the pure-$B_\varphi$ disk considered prior to the onset of the instability), the magnitude of the torque for the hybrid field configuration remains smaller than the HD value. However, in contrast with the 2D case (see Figure~\ref{fig:2DFFTorq}), it is larger (although not by much) than the magnitude of the pure-$B_\varphi$ torque. Since the magnitudes of the 3D torque contributions from the inner and outer disk regions are both larger in the pure-$B_\varphi$ model, the dominance in 3D of the net torque for the hybrid-field configuration is a consequence of a shift in the balance between these two contributions on going from 2D to 3D. The origin of this shift may be related to the apparent asymmetry in the radial extent of the innermost evanescent region between the inner and outer disk regions in the 3D hybrid-field case, particularly for low values of the azimuthal mode number $m$ (see lower panel of Figure~\ref{3DTPpanel}). Note that our turning-point analysis in Section~\ref{TPs}  also indicates that the size of the innermost evanescent region in the $m-r$ plane is smaller, in both the inner and outer disk regions, for the 3D pure-$B_\varphi$ model (upper panel of Figure~\ref{3DTPpanel}), which is consistent with our finding that the contribution to the torque from each of these regions is larger in this case than in the hybrid-field model.

A notable difference between the cumulative  3D torque for this field configuration and the 2D torque presented in Figure~\ref{fig:2DFFTorq} is that in 3D the total torque on the planet is negative, corresponding to inward migration, whereas in 2D it is positive (outward migration). A similar difference was already seen in the behavior of the pure-$B_\varphi$ field (again considering the situation in the 3D case before the onset of an instability). To understand the origin of this difference, we examine the contributions to the torque from different regions of the disk based on our linear analysis results. Table~\ref{tab1} shows the  torque exerted on the planet --- for particular azimuthal and (in 3D) vertical mode numbers --- by regions outside the LRs and by regions surrounding the magnetic and Alfv\'en resonances for both 2D and 3D realizations of this model. 

In the case of a pure-$B_\varphi$ field in 2D, \citet{Terquem03} found that the dominant contribution to the torque comes from the regions around the MRs even though the torque from the point-like contribution \emph{at} the resonances can sometimes be of opposite sign. We have confirmed this result and found that it applies also to the other cases considered in Table~\ref{tab1}: The point-like contribution from the MR can be opposite of that from the surrounding region for $m \lesssim 40$ (although not for higher values of $m$), and the overall MR contribution dominates the total torque. In 2D, the positive contribution from the inner MR region has a larger magnitudes than the negative contribution from the outer MR region for both the pure-$B_\varphi$ and the $B_z+B_\varphi$ field configurations, resulting in a positive net torque on the planet.\footnote{The $T_{10}$ torque value given in Table~\ref{tab1} for the 2D pure-$B_\varphi$ disk model (73 in the adopted units) can be directly compared with the corresponding entry in table~1 of \citet{Terquem03}, which is 66 in the same units (taking account of the opposite sign convention employed in that table). The numerical difference between these two values reflects the difference in the adopted radial range of integration ($0.3-1.7\,r_{\rm p}$ in our semianalytic calculations vs. $0.5-1.5\,r_{\rm p}$ in \citealt{Terquem03}; we reproduce the value in table~1 of the latter reference when we integrate over the narrower range).}  

For the 3D hybrid case, this trend is reversed: The outer MR region dominates the inner one, resulting in a negative net torque. This behavior can be understood from a consideration of the MR locations, using Equations \eqref{2Dres} and~\eqref{MRFull} for the 2D and 3D cases, respectively. Whereas the outer MR in the force-free case always lies farther away from the planet than the inner MR, in 2D the distances of the inner and outer MRs from $r_{\rm p}$ are comparable (their difference is $\sim 1\%$ of their mean). When $k_{z} \ne 0$, the asymmetry between the distance of the outer MR and inner MR is increased (the difference is $\sim 18\%$ of the mean for $k_zh=1.56$). This leads to the point-like contribution at the outer MR being significantly weaker relative to its surrounding region in the 3D case than in 2D. Hence, in 3D, the outer point-like contribution subtracts less from the torque arising from the surrounding region, resulting in that region's contribution dominating the net torque. As is seen from Table~\ref{tab1}, the effect of the MR regions in 3D is reinforced by that of the AR regions, which also contribute a negative torque from the outer side that outweighs the positive contribution of the inner side. As a check on these inferences, we decomposed the numerically derived density for the hybrid-field case into Fourier modes in the vertical direction, using $k_z = 2\pi n_z/L_z$ (where $L_z$ is the vertical domain size and $n_z$ is a nonnegative integer). We determined that the sum over the $n_{z}>0$ modes dominates the $n_{z}=0$ amplitude in the region around the outer MR. This is consistent with the linear-analysis arguments given above and supports the conclusion that the $k_z\ne 0$ modes are responsible for the reversal of the outward migration indicated in the 2D limit.

Unlike the pure-$B_\varphi$ disk, our hybrid field model (characterized by equal amplitudes of the azimuthal and vertical equilibrium field components) remains stable in 3D. We discuss the reason for the difference between these two cases in Section~\ref{discuss}.

\begin{table*}
\begin{center}
\caption{Torques from the Different Regions of a Force-Free, Uniform Disk in 2D and 3D\label{tab1}}
\begin{tabular}{@{}cccccc@{}}\toprule
\midrule
  \multicolumn{6}{c}{2D (for $m=10$)} \\
\midrule
 \textbf{B}  & $T_{Li}$ & $T_{MRi}$ & $T_{MRo}$  & $T_{Lo}$ & $T_{\rm total}$ \\ [0.5ex] 
\midrule
$B_{z} + B_{\varphi}$ & 39 & 257 & -233 & -45 & 18  \\
$B_{\varphi}$ &76 & 507 & -416 & -95 & 72 \\
\end{tabular}
\end{center}

\begin{center}
\begin{tabular}{@{}ccccccccc@{}}\toprule
\midrule
  \multicolumn{8}{c}{3D (for $m=10$ and $k_{z}h=1.56$)} \\
\midrule
 \textbf{B} & $T_{Li}$ & $T_{ARi}$ & $T_{MRi}$  & $T_{MRo}$ & $T_{ARo}$ & $T_{Lo}$  & $T_{\rm total}$\\ [0.5ex] 
\midrule
$B_{z} + B_{\varphi}$ & 1 & 114 & 316 & -422 & -134 & -1 & -126  \\
\midrule
\end{tabular}
\end{center}
\tablecomments{In 2D, the torque regions are defined as follows: $T_{Lo}$ is the torque from the region extending from the outer edge of the disk to $R_{L+}$, and $T_{MRo}$ is the region surrounding the MR from $R_{M+}$ to $r_{\rm p}$. In 3D, $T_{Lo}$  and $T_{MRo}$ enclose $T_{ARo}$ , which we define as the region surrounding the outer AR going from $R_{A+}$ to $R_{A-}$. See Figures~\ref{2DTPpanel} and~ \ref{3DTPpanel} for the turning-point labels. The inner regions (subscript `i' instead of `o'), are defined analogously.}
\end{table*}

\section{DISCUSSION}\label{discuss}

Our ultimate goal is to model planet migration in a generic wind-driving disk where the magnetic field is everywhere well-coupled to the matter. The midplane magnetic field configuration of such a disk is characterized by $B_z\ne0$, $B_\varphi=B_r=0$, $\partial B_\varphi/\partial z < 0$, $\partial B_r/\partial z > 0$, whereas near the surface the amplitudes of all three field components are comparable (with $|B_r|$, however, typically exceeding $|B_\varphi |$). A defining property of such a disk is that the magnetic torque term $\propto B_z \partial B_\varphi/\partial z$ accounts for a significant fraction --- and possibly all --- of the angular momentum transport that enables local gas to flow toward (and ultimately accrete onto) the central star. This transport is vertical, and in the case of a wind-driving disk the angular momentum is deposited in an outflow driven from the disk surfaces. As we have, however, pointed out, this model cannot be studied numerically with an ideal-MHD code, and a linearization treatment is similarly hampered by the need to incorporate magnetic diffusivity in order to establish an equilibrium field configuration in the differentially rotating disk. For this reason, we have concentrated in this paper on model problems that can be investigated within the framework of ideal MHD. While these problems do not directly mimic the expected situation in a wind-driving disk, they can increase our understanding of the possible effects of a large-scale, ordered magnetic field on Type~I planet migration. It is also conceivable that some of the field configurations that we explore might actually apply to real systems in certain circumstances.

The models that we considered --- pure-$B_z$, pure-$B_\varphi$, and a hybrid $B_z+B_\phi$ configuration, in either a force-free or a self-similar disk --- were examined in the 2D ($k_z \rightarrow 0$) and strict-2D ($v_z^\prime \rightarrow 0$) limits (with the latter case corresponding to the 2D implementation in numerical simulations) as well as in 3D using both semianalytic and numerical techniques. We now discuss the main insights provided by this study, which include the effect of a vertical field component, the difference between 2D and 3D models, and the onset of a disk instability in 3D.

As was pointed out by \citet{MutoEtal08} and confirmed in our work (see Sections~\ref{TPs} and~\ref{2Dsims_bz}, a purely vertical field in 2D simply acts to increase the effective sound speed in the disk, which pushes the turning points farther away from the planet and enhances the sharp, pressure-induced cutoff in the torque, thereby reducing the magnitude of the net torque acting on the planet in comparison with the unmagnetized disk case. We found that this is also the main effect of adding a $B_z$ component to a disk in which an azimuthal field is present (Section~\ref{2Dsims_combined}): the magnitude of the net torque is reduced but its sign (positive or negative) remains unchanged. Thus, in a force-free disk ($B_{\varphi}(r) \propto r^{-1}$) for which outward migration is predicted, the addition of a uniform vertical field $B_z = |B_\varphi(r_{\rm p})|$ reduces the cumulative torque by $\sim 35\%$ but keeps it positive (Section~\ref{forcefreesection}), whereas in a self-similar disk ($B_\varphi \propto r^{-5/4}$), for which inward migration is inferred, the addition of an equal-amplitude  vertical field component reduced the magnitude of the net torque by $\sim 50\%$ but keeps it negative (Section~\ref{selfsimilarsection}). This behavior can be understood from the fact that a vertical field component does not excite any new resonances in a 2D disk: its only impact is through the increase in the effective sound speed. The consequences of the increase in $c_{\rm eff}$ depend, however, on the nature of the disk model: For example, in a 2D model the addition of a vertical component to an azimuthal field shifts the MRs toward the planet, whereas in a strict-2D model such an addition pushes them away (Section~\ref{pure2Dsec}).

In contrast to its 2D behavior, a $B_z$ field in 3D gives rise to resonances --- both MRs and ARs (with the former being stronger than the latter), resulting in the vertical propagation of SMS and Alfv\'en waves and the establishment of weak magnetic wakes (although the Lindblad wake, associated with the propagation of FMS waves that are launched at the effective Lindblad resonances, still dominates the density perturbation; see Figure~\ref{fig:rho_bz_3d}). The opening up of additional wave propagation channels can be expected to increase the torque on the planet, yet we find (Section~\ref{3dsims_bz}) that the net torque is, in fact, reduced (by $\sim 50\%$ relative to the HD case for a uniform disk with $\beta_z=1$). Based on the results of a linear analysis (Section~II.4.2), we attribute this outcome to the fact that in this case the 2D mode dominates the contributions of the $k_z\ne 0$ vertical modes to the net torque. 

As we have seen, a pure-$B_\varphi$ disk that harbors a planet becomes rapidly unstable 
in 3D, but the addition of a vertical field component of comparable strength stabilizes it. The hybrid field configuration also develops (dominant) MRs and (subdominant) ARs, with the SMS and Alfv\'en waves now propagating along the \emph{total} field. Assuming a force-free field in a uniform disk, we find (Section~\ref{3dsims_bzbphi}) that the net torque in this case is negative --- the converse of the situation in the strict-2D limit --- which we argue is a consequence of a $k_z$-dependent asymmetry between the locations of the inner and outer MRs. It is noteworthy that the 3D torque for a force-free pure-$B_\varphi$ field (measured before the disk becomes unstable) is also negative, and that this is again the converse of the situation for the corresponding 2D case. This suggests that, in 3D, the role of the $B_z$ component in the hybrid field configuration is secondary to that of the $B_\varphi$ component, just as it evidently is in 2D. It is also worth noting that the magnitude of the 3D hybrid-field torque, while remaining smaller than in an unmagnetized disk, exceeds the magnitude of the (pre-instability) pure-$B_\varphi$ torque, which contrasts with the situation in the 2D hybrid-field models (Section~\ref{2Dsims_combined}). This difference might also be the result of a subtle effect such as turning-point asymmetry, and it provides another illustration of the fact that the details of the behavior in 3D are in general more complex (and less predictable in advance) than in 2D.

A potential caveat to our results is that the finite vertical domain of the simulations constrains the modes that can be resolved numerically: only modes with wavenumbers $k_{z}>k_{z,\rm min}=2\pi/L_{z}=
24.17$ (for $L_{z}=0.26$) can play a role in the simulation. This corresponds to $k_{z,\rm min}h=2.42$. In the case of unstratified disk models, $L_z$ is the only lengthscale that influences the relative importance of the vertical modes, which in principle can lead to an unphysical dependence of the simulation results on the domain length. We have verified that this is not a significant effect in our simulations by running vertically stratified disk models for the uniform-$B_\varphi$ and $B_\varphi(z)\propto z$ field configurations and determining that the calculated torques are very similar to those obtained for the unstratified disk models. We also doubled the vertical domain size for the simulated uniform hybrid-field model and confirmed that the derived torque remained essentially the same.

The pure-$B_\varphi$ field configuration considered in Section~\ref{3dsims_bphi} exhibits an instability that develops at the position of the planet and, within a few orbital periods, results in strong turbulence that continues to spread into the surrounding disk. The characteristics of this instability --- including the growth of the turbulent Reynolds and Maxwell stresses on a timescale on the order of $\Omega^{-1}$, with the stresses remaining comparable (yet dominated by the magnetic contribution) and saturating to a total-stress-to-gas-pressure-ratio of $\alpha \gtrsim 10^{-2}$, as well as the insensitivity to background stratification --- have the hallmarks of a ``standard'' MRI , which arises in a Keplerian disk that is threaded by a weak vertical field \citep[e.g.,][]{BalbusHawley98}. This interpretation may at first  seem  puzzling since a similar instability does \emph{not} develop for either the pure-$B_z$ or the hybrid ($B_{\varphi}+B_z$) field configuration. The  absence of this instability in the latter two models can, however, be understood from the existence of a minimum vertical wavelength for the development of a standard MRI (arising from the need to overcome the stabilizing effect of the magnetic tension force) and from the requirement that this wavelength remain less than the disk thickness. This condition can be expressed as a lower limit on the parameter $\beta_z$: $\beta_{z,\rm min}=(\pi/\sqrt{3})(h/r_0)$ for instability. For $h_{\rm p}=0.1$ and $z_0=0.13$, the values used in our simulations, we obtain $\beta_{z,\rm min} = 1.4$. Since our pure-$B_z$ model employs $\beta_{z\rm p} =1$, and so does also our hybrid model, both are stable to the standard MRI.

A pure-$B_\varphi$ disk is also subject to a weak-field shearing instability, although, unlike the standard MRI, it is triggered by the excitation of nonaxisymmetric, rather than axisymmetric, modes \citep[e.g.,][]{HawleyEtal95,RuedigerEtal07}. In the limit of large azimuthal wave numbers ($m \gtrsim h^{-1}$), one can write down the instability condition \citep[e.g.,][]{OgilviePringle96,TerquemPapaloizou96} and again convert it into a lower bound on the relevant $\beta$ parameter: $\beta_\varphi > \beta_{\varphi,\rm min} = m h/\sqrt{3}$ for instability. The numerical value of $\beta_{\varphi,\rm min}$ should be comparable to that of $\beta_{z,\rm min}$, and one can therefore expect that our pure-$B_\varphi$ disk model, with $\beta_{\varphi \rm p}=1$, would also be stable to the MRI (the \emph{azimuthal} MRI (AMRI) in this instance). In fact, even in the case of instability, the growth rate of the AMRI would be much slower than that of the standard MRI (i.e., $\ll \Omega^{-1}$) unless $k_z h$ were $\gg 1$, in which case the presence of even a tiny poloidal field component could transform the nature of the problem \citep[see][]{BalbusHawley98}. It seems that this is precisely what has transpired in our 3D pure-$B_\varphi$ simulation: This configuration evidently remains stable to the AMRI, but \emph{the growth of the perturbed vertical field component (of initial amplitude $\lesssim 10^{-3}|B_{0\varphi}|$), induced by the gravitational perturbation of the planet}, triggers the standard MRI and turns the disk turbulent over a few orbital periods. We note in this connection that \citet{HawleyEtal95} previously demonstrated that the MRI behavior of a hybrid $B_\varphi + B_z$ field configuration is roughly the same as that of the sum of the two individual models; in particular, they found that the evolution of a standard MRI triggered by a weak vertical field is unaffected by the presence of a comparatively strong (AMRI-stable) azimuthal field in the disk.

The field configurations that we considered are potentially subject also to other types of instabilities. For example, the hybrid case was shown by \citet{HollerbachRuediger05} to be susceptible to the development of a nonstationary flow pattern, consisting of elongated Taylor vortices that drift in the vertical direction. This instability can be attributed to the handedness of this field configuration, which is only present when both the $B_z$ and $B_\varphi$ components are nonzero. In our simulation of the hybrid field model we also observe the development of a nonsteady flow pattern that breaks the $\pm z$ symmetry, but a longer run that covers a larger domain is needed to study this behavior in full detail.

\section{CONCLUSION}\label{conclude}

We have studied Type I migration in protoplanetary disks that are threaded by a large-scale, ordered magnetic field. Among the possible origins of such a field in real disks are interstellar field lines that have been carried in by the accretion flow and a dynamo-generated field from the central star that has diffused into the disk gas. In either of these two cases, the field can be modeled as being open and possessing an even symmetry ($B_r=B_\varphi = 0$, $B_z \ne 0$ at the midplane). In general, the drag exerted by the inflowing gas and the differential rotation of the disk give rise to vertical field gradients --- ($\partial B_r/\partial z$ and $\partial B_\varphi/\partial z$ --- that, in conjunction with the vertical field component $B_z$, produce a radial magnetic tension force and a braking torque, respectively. The torque term can lead to significant vertical transport of angular momentum, which, in turn, may be deposited in a powerful wind that is driven centrifugally from the disk surfaces. Strong outflows of this type are a ubiquitous feature of protoplanetary systems, and it is of interest to explore what effect the magnetic angular momentum transport associated with them might have on planet migration in the wind launching region. However, the general problem cannot be properly studied within the framework of ideal MHD because the strong winding up of the field lines in a Keplerian disk leads, in the absence of magnetic diffusivity, to an ever growing torque that results in an unchecked growth of the inflow velocity. A further complication is associated with the fact that a wind-driving disk cannot be treated as a spatially localized problem. For these reasons we have chosen to concentrate in this paper on simplified field configurations that can be handled in a restricted domain and using ideal MHD, in the hope that they could provide useful insights into the more general problem. Some of these field geometries might even be relevant to real systems, and, in any case, their treatment enables us to make contact with (and generalize) previous studies in the literature. 

The field geometries that we considered are: pure-$B_\varphi$ (previously studied semianalytically and numerically in the 2D limit by \citealt{Terquem03} and \citealt{FromangEtal05}, respectively), pure-$B_z$ (previously considered semianalytically and numerically, but exclusively in the shearing-sheet approximation and employing additional simplifications, by \citealt{MutoEtal08}), and a hybrid ($B_\varphi + B_z$) field configuration, where the fiducial $\beta$ parameter value for each of the field components was, in all instances, 1. In the case of the pure-$B_\varphi$ field, we studied both a vertically uniform configuration and a model in which the field grows linearly with height from zero at the midplane.  We examined uniform, force-free disks, in which neither a thermal nor a magnetic force plays any role in the equilibrium state, as well as disks in which the radial dependence of the physical variables corresponds to that of the radially self-similar model of wind-driving disks. Our study comprised a linear perturbation analysis and numerical simulations in both 2D and 3D, where we distinguished between the analytic 2D limit of setting the vertical wavenumber $k_z$ equal to zero and the stricter numerical limit of letting the vertical component of the perturbed velocity vanish. Our linearization analysis has yielded the relevant resonances for the different field configurations and disk geometries and the associated wave propagation regions. We identified magnetic and Alfv\'en resonances and obtained their locations with respect to the corotation radius, and we presented algebraic expressions for these quantities (derived under the assumption of $z$ invariance).

For any given model, we derived semianalytically the turning-point locations that delineate the propagation regions of the respective waves (slow-magnetosonic and Alfv\'en) as well as the locations of the magnetically modified effective Lindblad resonances from which fast-magnetosonic waves are launched. We then evaluated the torques contributed by different regions in the disk and attempted to estimate the magnitude of the cumulative torque for that case. This information was used to analyze and complement the density maps and the torque running-time averages obtained in the numerical simulations. The main findings of this study can be summarized as follows: 

\begin{itemize}

\item The inference from 2D studies of pure-$B_\varphi$ disk models that a sufficiently strong outward decrease in the magnetic field amplitude would lead to outward planet migration does not apply to the 3D models that describe real systems. In 3D, planet migration is invariably inward for each of the field configurations and disk models that we considered.

\item In the presence of a planet, a disk threaded by a uniform azimuthal field that is characterized by a field-strength parameter $\beta_{\varphi \rm p}$ large enough to forestall the development of the azimuthal MRI (AMRI) becomes unstable at the planet's location to the standard MRI (which is associated with the presence of a weak vertical magnetic field) and turns turbulent within a few orbital periods. This outcome is found irrespective of whether the effect of the vertical component of the stellar gravity is taken into account. The instability is triggered by the growth of the perturbed vertical magnetic field component that is induced by the disk's interaction with the planet. This instability did not develop in either the pure-$B_z$ or the hybrid field model since our chosen value of $\beta_z$ was large enough to forestall the standard MRI in both of these cases.

\item Even though the presence of a large-scale field cannot (in 3D) reverse the inward migration predicted by the underlying unmagnetized disk model, it generally slows it down --- by a factor of $\sim 2$ for a pure-$B_z$ configuration and a somewhat lower value
for the hybrid-field case (using $\beta_{z\rm p}=\beta_{\varphi\rm p} = 1$).

\end{itemize}

In the companion paper we carry out a linear perturbation analysis of a disk model that incorporates the vertical magnetic angular transport term ($\propto B_z \partial B_\varphi/\partial z$) in an attempt to get a more direct handle on the planet migration properties of wind-driving disks. We plan to eventually also perform a 3D nonideal-MHD simulation to gain even stronger insights into this problem.

\acknowledgments
We thank the referee, Takayuki Muto, for his insightful comments and
suggestions that greatly improved this manuscript. We also thank Fausto 
Cattaneo and Krista Martocci for their useful input and discussions. This research was supported in part by NSF grant AST-0908184 and NASA ATP grant NNX13AH56G, as well as by a NASA Earth and Space Science Fellowship Program grant NNX09AQ89H 
to A.B. Numerical simulations were carried out at the Midway High Performance Computing Cluster at the Research Computing Center of the University of Chicago.

\appendix
\section{Softening Parameter in 2D Simulations}
\label{appA}
It is well known that the derived magnitude of the
torque in 2D simulations depends on the choice of softening parameter $\epsilon$ in Equation
\eqref{eq:gpot}. For disk--planet interactions in unmagnetized disks, it has been
argued that the softening parameter should be about $0.6\, h$ in order for the
torque values to match the ones obtained in 3D simulations
\citep{PaardekooperPapaloizou09a,PaardekooperPapaloizou09b,PaardekooperEtal10}. In this work we do not use this scaling for the following reasons. First, we do not know
whether this particular choice of $\epsilon$ is also applicable to magnetized
disks. Second, the magnetic resonances that play a key role in 2D magnetized disk models generally lie closer to the planet than the Lindblad resonances that dominate in the unmagnetized case, which adds the constraint that the distance modeled by this parameter should be less than the radial distance from the planet to the MRs. Third, we do not attempt to match the 2D and 3D torque values since we are actually interested in the difference between the 2D and 3D disk models: In the magnetized case, new resonances appear in 3D, along with new regions and directions of wave propagation, making the 3D problem
fundamentally different from the 2D one. Now, the 2D Lindblad and corotation contributions, which are the relevant torque term in the HD case, were shown to converge to a given value when $\epsilon$ is small \citep{PaardekooperPapaloizou09b}, so we take this value to represent the 2D HD torque in the present study. Motivated by these considerations, we fix the magnitude of $\epsilon$ at 0.03, which is small in comparison with $h$ but still large enough to 
prevent the occurrence of numerical divergences in the vicinity of the planet.

\section{Initializing the Disk Density Profile in the Presence of Vertical Stellar Gravity}
\label{appB}
We derive an approximate solution for the vertical density structure of a disk that is subject to compression by the tidal gravitational field of the central star as well as by the magnetic pressure gradient associated with the azimuthal field component. This solution is used to initialize the simulations presented in Figure~\ref{fig:ver_struc_bphi_zgrad} in order to avoid the development of artificial currents. For a vertically isothermal disk, the $z$ component of the momentm equation \eqref{eom} reads
\begin{equation}
c^{2}\frac{\partial \rho}{\partial z} = -\rho \frac{GM}{(r^{2} +
z^{2})^{3/2}}z - \frac{B_{\varphi}}{4\pi}\frac{\partial B_{\varphi}}{\partial z}\ .
\label{eq:vert_eq}
\end{equation}
Under our adopted assumption, $B_\varphi \propto z$ (see Section~\ref{3dsims_bphi_zgrad}), so we can write $B_\varphi = (\partial B_\varphi/\partial z) z$, with $\partial B_\varphi/\partial z=B_{\varphi \rm p}/z_0$ a constant. Assuming a geometrically thin disk ($z\ll r$) and focusing on the radial location of the planet, the approximate solution of Equation \eqref{eq:vert_eq} is
\begin{equation}
\rho_0(z) \approx \rho_{\rm p}\; exp\left\{-\left[1 + \frac{1}{\beta_{\varphi}}\left ( \frac{h_{\rm p}}{z_{0}} \right)^2 \right ]
\frac{z^{2}}{2h_{\rm p}^{2}} \right\}\ ,
\label{eq:rhoz}
\end{equation}
where $h_{\rm p}$ is the (normalized) tidal scale height (Section~\ref{num_setup}) and $\beta_\varphi$ is evaluated at the top of the disk (where $B_\varphi = B_{\varphi \rm p}$) but using the midplane density $\rho_{\rm p}$ (an approximation justified by the fact that the density does not change much over either $h_{\rm p}$ or $z_0$ for the adopted values of these parameters ).

%\begin{thebibliography}{}
%\end{thebibliography}

\bibliographystyle{apj}
\bibliography{mybib1}
\clearpage

\end{document}